\documentclass[twocolumn]{aa} 
\usepackage{graphicx}
\usepackage{ gensymb }
\usepackage{ amssymb }
\usepackage{ amsmath }
\usepackage{ tipa }
\usepackage{ textcomp }
\usepackage{ wasysym }
\usepackage{float}
\usepackage{graphicx,color}
\usepackage{xcolor}
\newcommand{\jm}[1]{#1}

\usepackage{rotating}
\usepackage{txfonts}
\usepackage{lscape}
\begin{document}

   \title{\jm{The role} of atom tunneling in \jm{gas-phase} reactions in \jm{planet-forming} disks}

 \author{
    J. Meisner\inst{1,2}
        \and
    I. Kamp \inst{3}
        \and
        W.-F. Thi\inst{4}
    \and
        J. K\"{a}stner\inst{1}
        }

        \institute{Institute for theoretical chemistry, University of Stuttgart, Germany
        \and Current Address: Department of Chemistry and The PULSE Institute, Stanford University, Stanford, California 94305, USA\\ 
        \email{meisner@theochem.uni-stuttgart.de}
    \and Kapteyn Astronomical Institute, University of Groningen, The Netherlands\\ 
        \email{kamp@astro.rug.nl}
        \and
     Max Planck Institute for Extraterrestrial Physics, Gie\ss enbachstrasse 1, 85741 Garching, Germany}
 

 
  \abstract
 {Gas-phase chemical reactions of simple molecules have been recently revised to include atom tunneling at very low temperatures. This paper investigates the impact of the increased reaction rate constant due to  tunneling effects on \jm{planet-forming} disks.}
  {Our aim is to quantify the astrophysical implications of atom tunneling for simple molecules that are frequently used to infer disk structure information or to define the initial conditions for planet (atmosphere) formation.}
   {We quantify the tunneling effect on reaction rate constants by using H$_2$~+~OH~$\rightarrow$~H$_2$O~+~H as a scholarly example in comparison to previous UMIST2012 rate constants. In a chemical network with 1299 reactions, we identify all chemical reactions that could show tunneling effects. We devise a simple formulation of
   \jm{reaction rate constants} that overestimates tunneling and screen a standard T~Tauri disk model for changes in species abundances. 
For those reactions found to be relevant, we find 
\jm{values of the most recent literature}
for the rate constants including tunneling and compare the resulting disk chemistry to the standard disk model(s), a T~Tauri and a Herbig disk.}
   {The rate constants in the UMIST2012 database  in many cases already capture  tunneling effects implicitly, as seen in the curvature of the Arrhenius plots of some reactions at low temperature. A rigorous screening procedure identified three neutral-neutral reactions where atom tunneling could change simple molecule abundances. 
However, by adopting recent values of the rate constants of these reactions and due to the layered structure of \jm{planet-forming} disks, the effects are limited to a small region between the ion-molecule dominated regime and the ice reservoirs where cold ($<250$~K) neutral-neutral chemistry dominates. 
Abundances of water close to the midplane snowline can increase by  a factor of  two at most compared to previous results with UMIST2012 rates. Observables from the disk surface, such as high excitation ($>500$~K) water line fluxes, decrease by  60\% at most when tunneling effects are explicitly excluded. On the other hand, disk midplane quantities relevant for planet formation such as the C-to-O ratio and also the ice-to-rock ratio are clearly affected by these \jm{gas-phase} tunneling effects.}
{}

    \keywords{astrochemistry; 
    Protoplanetary disks; 
    Stars: low-mass; 
    Infrared: planetary systems}
           \maketitle
%

%
%
%

\section{Introduction}

Observations and modeling of molecules are relevant in order to obtain a chemical inventory in space and to derive physical conditions in astrophysical environments. For  example, the HEXOS program, a Herschel/HIFI key program  \citep{Bergin2010},  revealed the chemical inventory of the Orion and Sagittarius B2 star forming regions. The team identified complex organic species such as methanol (CH$_3$OH), hydrogen cyanide (HCN), and methyl cyanide (CH$_3$CN),
among others, which are thought to be precursors of amino acids \citep{Crockett2014}. On the other hand, we can use the line emission of more simple species to learn about the physical conditions in space, for example  the fine structure and molecular lines observed towards the dark cloud core Barnard 5 \citep{Bensch2006}. 
Comparing the line ratio [C\,{\sc i}]/CO J=3-2 with 
\jm{models of a 1D spherical grid of the photodissociation region 
can give us an estimation of} the total mass and density of the region close to the central core. Other examples are interferometric measurements of CO submm lines in protoplanetary disks that yield the radial extent of the molecular gas disk \citep{Isella2007}, and CN lines that can probe the vertical structure  (e.g.,\ flaring and UV penetration) of \jm{planet-forming} disks \citep{Cazzoletti2018,vanZadelhoff2003}.

To assist the quantitative interpretation of such astrophysical observations, extensive chemical modeling is often performed using large databases of chemical reaction rates such as \jm{the UMIST Database for Astrochemistry 2012 (UMIST2012)} \citep{McElroy2013b}, \jm{the Kinetic Database for Astronomy (KIDA)} \citep{Wakelam2012}, 
or 
\jm{the Ohio State University (OSU) database} \citep[e.g.,][]{Harada2010}. 
The data in these first two compilations are regularly reviewed and updated. \jm{However, these data are often} based on experimental work carried out in a 
finite
temperature range. 
Cryogenic temperatures 
are often hard to access in the laboratory as the rate constants become too small to be measured precisely 
as reactants freeze out and unwanted reactions with surfaces of the experimental setting take place.
Hence, computational methods gain increasing importance when trying to accurately assess chemical reactivity \citep{meisnerreview2016,biczysko2018}.
Despite explicit warnings \citep{Woodall2007,Wakelam2012}
extrapolation to lower temperatures is performed frequently,
often using Arrhenius-type formulas.

In the last decades, the accuracy of rate constants obtained by means of computational methods
has increased significantly, in particular at low temperatures where quantum mechanical effects can become relevant \citep[see][for a recent white paper on these advancements]{Wiesenfeld2016}.
One nuclear quantum effect we want to address in this paper is the tunneling effect, also referred to as quantum tunneling.
Caused by the wave particle dualism, the quantum mechanical tunneling describes the effect that a particle can penetrate a potential energy barrier the particle could classically not overcome.
In general, quantum tunneling plays a role in various fields of physics.
In chemistry, the tunneling of atoms through a potential energy  barrier 
can allow reactivity even in cases where it 
would classically be forbidden.
Examples can be found in biochemistry, molecular  catalysis, and  surface science 
\citep[see][for a recent review]{meisnerreview2016}.
For an overview of the importance of atom tunneling for 
astrochemical surface reactions, we refer to the review by \citet{hama2013}.


The  tunnel effect contributes considerably to 
the rate constants of chemical reactions, both in the gas phase 
and on ice surfaces.
 Prime examples for surface reactions enhanced by atom tunneling are
 hydrogen atom additions to unsaturated chemical compounds.
 The subsequent hydrogenation of CO to formaldehyde and finally to methanol \citep{hidaka2009,andersson2011,goumans2011b} is one of the most prominent
 examples.
The addition of hydrogen atoms to graphite surfaces or polycyclic aromatic 
hydrocarbons (PAHs) is assumed to be an intermediate step in the formation of  H$_2$  \citep{goumans2010,wakelam2017}. 
In addition to these instances, water formation was shown to be 
enhanced by atom tunneling  on ice surfaces by \citet{oba2012}.

In the gas phase, atom tunneling can play an important role in the formation of complex organic molecules (COMs) and can, for example, change the
results of  kinetic models qualitatively.
For instance, the rate constants of some cryogenic 
reactions even increase with further lowering the temperature, 
\jm{see below} \citep{shannon2013,alvarez-barcia2016}.
As light $^1$H (protium) atoms are more likely to undergo the tunneling 
process than $^2$D (deuterium) atoms, hydrogen abstraction of COMs 
is more likely than deuterium abstraction, for example leading to a deuterium enrichment 
of  methanol \citep{goumans2011a}.


The tunneling probability of a particle depends on the width of the barrier, on the mass of the tunneling particle,  and 
on the height of the potential energy barrier. 
A suitable estimate on the relevance of atom tunneling for a particular reaction is the crossover temperature $T_{\text{C}}$,
which can be obtained easily by current computational methods.
The crossover temperature depends on the shape of the potential energy barrier, more precisely on the negative curvature along the
reaction path in mass-weighted Cartesian coordinates, $\lambda_i$, 
\begin{equation}
T_{\text{C}}=   \frac{\hbar \sqrt{\lambda_i}}{2\pi k_{\text{B}}} ,
\end{equation}
where $k_{\text{B}}$ is Boltzmann's constant and $\hbar$ is Planck's constant,
The crossover temperature can be understood as the temperature below which atom tunneling is more likely than the classical transition.
For most  hydrogen atom or proton transfer reactions, the
crossover temperature is between 100~K and 300~K, 
while $T_{\text{C}}$ is rarely above 150~K when the motion of heavier atoms is involved.

\begin{figure}[htb]
\begin{center}
  \includegraphics[width=8cm]{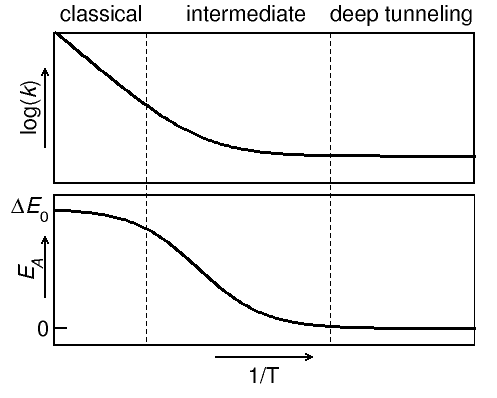}
\end{center}
\caption{
Top: Schematic Arrhenius plot in the different temperature regimes for a unimolecular reaction. 
Bottom: Corresponding activation barrier $E_A$ (modified from \citealt{meisnerthesis2018}).
\label{fig:arrh1}
}
\end{figure}

Classically, the reaction rate constant decreases with decreasing temperature
for an elementary reaction.
When the classical 
over-the-barrier reaction
dominates, the Arrhenius plot, which is the plot of the logarithmic reaction rate constant against the inverse of the temperature, is expected to decrease linearly (see Fig.~\ref{fig:arrh1}).
However, at lower temperatures the Arrhenius plot becomes curved due to 
a non-negligible contribution of atom tunneling \citep{meisnerreview2016}.
 At very low temperatures, tunneling can occur from a single quantum state
leading to a temperature-independent rate constant \citep{zuev2003}.

The slope of the Arrhenius plot is assigned to be proportional to the 
activation barrier of a chemical reaction $E_A$:
\begin{equation}
E_A = - R \, \frac{\mathrm{d} \ln(k(T))}{\mathrm{d} \, 1/T }.
\end{equation}
Following Tolman, the activation energy can also be understood as the average energy of the molecules reacting minus the average energy of all reactant molecules \citep{tolman1920,truhlar1978}.
The activation energy in general depends on the temperature and in nearly all cases 
decreases towards lower temperatures, implying
a positive curvature, i.e.,
\begin{equation}
\frac{\mathrm{d}^2 \ln(k(T))}{\left(\mathrm{d}~1/T \right)^2 } > 0,
\end{equation}
even though some convex exceptions exist, as described in \citet{truhlar2000}.

The activation energy, however, is not directly related to the 
potential energy barrier  $\Delta V_0$ 
or the adiabatic reaction barrier $\Delta E_0$
(which is the zero-point corrected potential energy barrier)
as prominent examples will demonstrate later.
At higher temperatures, when there is enough energy for the particles to
overcome the potential energy barrier and react classically, 
the activation barrier can be quite close to  the adiabatic reaction barrier $\Delta E_0$.
 When nuclear quantum effects are more important (e.g., at lower temperatures) this is no longer the case. 
As described above, the temperature regimes differ from reaction to reaction \citep{meisnerreview2016}.

In astrochemistry, the modified Arrhenius equation \jm{(Parenthesis now added)} 
\begin{equation}
k(T)=\alpha \left( \frac{T}{300~{\rm K}} \right)^{\beta} \exp{\left(-\frac{\gamma}{T}\right)}
\label{eq:modArrh}
\end{equation}
is used for chemical modeling. 
Here, the parameters 
$\alpha$, $\beta$, and $\gamma$ are obtained by fits to either 
experimental measurements or 
from accurate computations.
In contrast to the physical interpretations of $E_A$ and  
discussed above,  
$\alpha$, $\beta$, and $\gamma$ 
are merely fitting parameters without any physical meaning 
and are only applicable for the temperature regime the original data stem from, 
while extrapolation should in general be avoided.
Although Eq.~(\ref{eq:modArrh}) looks similar to the Eyring equation, 
effects which are not covered in classical Eyring theory are 
implicitly included in the data forming the basis for the fit, and thus 
are implicitly included in the fitting parameters $\alpha$, $\beta$, and $\gamma$.
In particular, the value of $\gamma$ is in most cases lower
than the adiabatic energy barrier when fitted at moderate temperatures, due to atom tunneling.

\label{sect:examplewaterreaction}

\begin{figure}[htb]
\includegraphics[width=9cm]{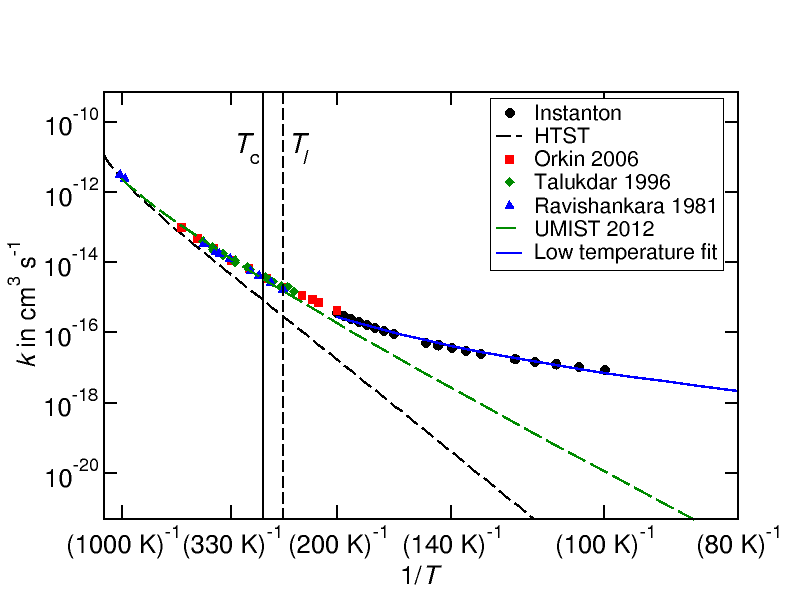}
\caption{Arrhenius plot for the  reaction H$_2$~+~OH~$~\rightarrow$~H$_2$O~+~H.
The low-temperature fit (blue solid line) was performed using the rate constants
calculated using the instanton theory by \citet{meisner2016}.
Experimental values are from 
\jm{
\citet{ravishankara1981}, 
\citet{talukdar1996}, 
and \citet{orkin2006}.}
     The vertical dashed line indicates the lower bound for the recommended temperature range of the UMIST2012 database,  $T_l =250$~K.
Reprinted (modified) with  permission from \emph{J. Chem. Phys.} \textbf{144}, 174303 (2016), Copyright (2016) American Institute of Physics.
}
\label{fig:H2+OH}
\end{figure}

A nice example where $\Delta E_0$ and $\gamma$ can  be accidentally  mixed up is the reaction
\begin{equation}
{\rm H_2 + OH}  \rightarrow {\rm H_2O + H}.  \hfill [1] \nonumber 
\end{equation} 
In this radical neutral-neutral reaction, a hydrogen atom is transferred from the hydrogen molecule forming water and a remaining H atom. 
For this reaction, atom tunneling sets in at comparably high temperatures;
the crossover temperature is approximately 276~K, indicating a rather sharp potential energy barrier \citep{meisner2016}.
However, atom tunneling affects the reaction rate constants at temperatures 
as high as 500~K, which can be recognized by the noticeable curvature in Fig.~\ref{fig:H2+OH} and 
 the deviation of the experimental values from the rate constants  calculated with  harmonic transition state theory (HTST), which  uses quantum mechanical expressions for the translational, rotational, and vibrational partition function (thus including  zero-point energy), but does not include atom tunneling. 
It is therefore perfectly suited to monitoring the influence of
atom tunneling on 
reaction rate constants.
Details of the calculations can be found in \citet{meisner2016}.

Due to the importance of atom tunneling for this reaction,
the experimentally observed activation energy of approximately
2000~K (depending on the temperature region where the values are taken) 
deviates significantly from the potential and adiabatic energy barriers of $\approx$~2700~K and $\approx$~2900~K, respectively.
This mix-up 
leads to some confusion about the actual barrier height \citep{meisner2017}.
In UMIST, $\gamma=1736$~K, implicitly covering the effect of atom tunneling, although underestimating it at cryogenic temperatures. 

Another example reaction where the value of $\gamma$ is not related at all to the
adiabatic energy barrier
is the hydrogen atom transfer reaction
NH$_3^+$~+~H$_2 \rightarrow$~NH$_4^+$~+~H, \citep{smith1991,herbst1998, alvarez-barcia2016},
which possesses a strongly stabilized pre-reactive complex 
leading to a positive slope of the Arrhenius plot at cryogenic temperatures \citep{shannon2013},
and therefore $\gamma=-50.9$~K in the UMIST2012 database.
However, the adiabatic energy barrier of this reaction is 
$\Delta E_0=1460$~K.

As discussed above, atom tunneling is an important feature of chemical 
kinetics in both gas-phase and solid-phase reactions. 
In this work, we focus on the atom tunneling effect of gas-phase 
reactions, the modified reaction rate constants, and the impact on 
protoplanetary disks. 
This is a good starting point to elucidate the relevance of atom tunneling
for a particular astrophysical environment being distinct from  dense clouds, for example. 
Furthermore, properly including atom tunneling in surface reactions 
would also need to take atom tunneling  into account for 
diffusion processes \citep{oba2012}, which  is clearly beyond the scope of this 
work.


In this work, the influence of the tunnel effect 
on gas-phase reactions in \jm{planet-forming} disks is studied systematically. This paper does not assess the quality of the UMIST2012 database or the accuracy of the disk models, but focuses on the influence of quantum tunneling on the whole kinetic model.
It is, to the best of our knowledge,
the first  study of the impact of the tunneling effect on a large chemical reaction network, rather than  atom tunneling in just one specific reaction.


The paper is structured as follows.
In   Sect. 2 
we describe the underlying disk models including the procedure used to select chemical reactions that could show tunneling effects. We also describe the methodology of our analysis.
In  \jm{Sect. 3} we present the influence of the quantum mechanical tunnel effect 
on the molecular composition of \jm{planet-forming} disks,  based on the systematically varied Arrhenius parameters 
for the different chemical reactions. We also present the  species densities 
and line fluxes.
In \jm{Sect. 4} we discuss the astrophysical implications of the 
  results reported here, and in \jm{Sect. 5} we present our conclusions.


\section{Methods}

To elucidate the importance of atom tunneling 
for \jm{gas-phase} chemistry in the environment of \jm{planet-forming} disks, we calculated reaction rate constants of the water forming reaction H$_2$~+~OH~$\rightarrow$~H$_2$O~+~H without  any atom tunneling and compare that to the UMIST2012 database \citep[see below for technical details]{McElroy2013b}.

In the recent literature, rate constants 
for some chemical reactions 
have been calculated using the semi-classical instanton theory.
These rate constants can be considered  reasonably accurate and sufficiently precise 
to be used in quantitative kinetic models.
We used rate constants 
for the reactions of 
H$_2$~+~OH~$\rightarrow$~H$_2$O~+~H
\citep{meisner2016},
CH$_4$~+~OH~$\rightarrow$~CH$_3$~+~H$_2$O
\citep{lamberts2017},
and
CH$_4$~+~H~$\rightarrow$~CH$_3$~+~H$_2$
\citep{beyer2016}.


More chemical reaction rate constants with comparable or even better accuracy may exist in the literature. However, 
a complete literature review of all chemical reactions is not feasible, and hence we perform here first a screening
%
to identify additional reactions where atom tunneling might be important.
We then perform an exhaustive screening of those reactions using \jm{planet-forming} disk models as astrophysical environments. For the reactions where the screening actually identifies large effects of tunneling on species abundances, recent literature values are looked up, compared to the UMIST2012 database, and used in a comparative disk simulation.


\subsection{No tunneling}
\label{sect:no-tunnel}

We used the  reaction rate constants of the reaction of 
\begin{equation}
{\rm H_2 + OH}  \rightarrow {\rm H_2O + H}  \hfill [1] \nonumber 
\end{equation}
calculated with HTST by  \citet{meisner2016}.
The fit of the modified Arrhenius equation (\ref{eq:modArrh})
gives 
\begin{equation}
k(T)=2.157 \cdot 10^{-11} \left( \frac{T}{300~{\rm K}} \right)^{0.2583} 
\exp{\left(-\frac{2791.0~\text{K}}{T}\right)}~~~\frac{\text{cm}^3}{\text{s}}
\end{equation}
for the reaction rate constants in absence of atom tunneling.
It should be noted that the value of $\gamma$~=~2791.0~K is 
closer to the adiabatic energy barrier of
2913~K \citep{meisner2016}
than to the UMIST2012 value of 1736~K.

At around 500~K, the  experimental rate constant is already a factor 
of 5 higher than the  rate constant obtained by HTST. 
At 200~K, this ratio increases to 13 at 200~K and is expected to be several orders of magnitude higher at $\approx 10$~K.  
The differences of the rate constants are shown in Fig.~\ref{fig:H2+OH}. 
The values of the HTST calculations are  from ~\citet{meisner2016}.




\subsection{Screening of chemical reactions that might be influenced by atom tunneling}
\subsubsection{Creation of the list of reactions}

We used the small chemical network with 100 species and 1299 reactions from \citet{Kamp2017} to identify chemical reactions that could show effects of H-tunneling using the following criteria:\\[-5mm]
\begin{itemize}
\item The reaction is   neutral-neutral or ion-neutral. 
Other types of reactions do not involve a simple ground-state over-the-barrier kinetics, and therefore make it difficult to distinguish
the tunnel effect from other nonclassical effects.\\[-3mm]
\item The reaction does not involve excited molecular hydrogen (H2exc) as a reactant or product. \\[-3mm]
\item $\gamma$~>~0~K, which implies that there is a potential energy barrier. 
As discussed above, the fitting parameter $\gamma$ does not directly relate to a barrier height. 
A negative  value of $\gamma$ leads to an increase in reaction rate constants at very low temperatures. This behavior can be observed for molecules with a 
strongly pronounced pre-reactive complex in combination with 
a significant contribution of atom tunneling, like that in the 
reaction of OH radicals with methanol (see \citealt{shannon2013}).
A simple modification like the one  we perform for this screening is therefore not sufficient to estimate  the rate constants at very low temperatures, and would even underestimate them.
%
Furthermore, the value of $\gamma$ is affected by atom tunneling, and
thus the biggest contribution 
is already incorporated in the fitting parameters provided by UMIST. This means that reactions with $\gamma~<~ 0 $ do not serve as a suitable test set for our study.
%
It is, however, strongly recommended to further investigate these reactions.
\\[-3mm]

\item $\gamma$~<~$10^{4}$~K, since these reactions can be assumed to posses relatively high potential energy barriers.
These reactions are instead related to the field of combustion chemistry
where the effect of atom tunneling is negligible due to the 
high temperature.
It can be assumed that atom tunneling has just a marginal influence on the reaction rate constants, if they are even measurable or computable.
However, high values of $\gamma$ might also occur in cases of endothermic reactions.
\\[-3mm]
\item The reaction transfers one single hydrogen atom or proton as 
the tunnel effect depends strongly on the mass of the transferred particles 
and hydrogen is the lightest chemical element.
The tunneling effect is therefore expected to have a 
larger impact than  heavier atoms have, even though  carbon
tunneling  also influences reactivity (\citet{zuev2003,borden2016})
\\[-3mm]
\item More than one product is formed. Otherwise, the excess energy resulting from the chemical reaction leads to the destruction of the product molecule.

\item
$\beta \ge 0$: a negative value of $\beta$, i.e., a negative curvature of the Arrhenius plot indicates that other dynamic effects dominate the reaction kinetics. Atom tunneling plays a minor role for these reactions.
\\[-5mm]
\end{itemize}
Applying these criteria to the rate network yields 47 reactions 
that perhaps show a significant speed-up by 
atom tunneling: 43 of these reactions are neutral-neutral reactions, 4 are of  ion-neutral. 
%
As the UMIST2012 data base is focused on gas-phase reactions, no surface reactions are considered. 

\subsubsection{Modification of the Arrhenius plots}
\begin{figure}[htb]
\includegraphics[width=9cm]{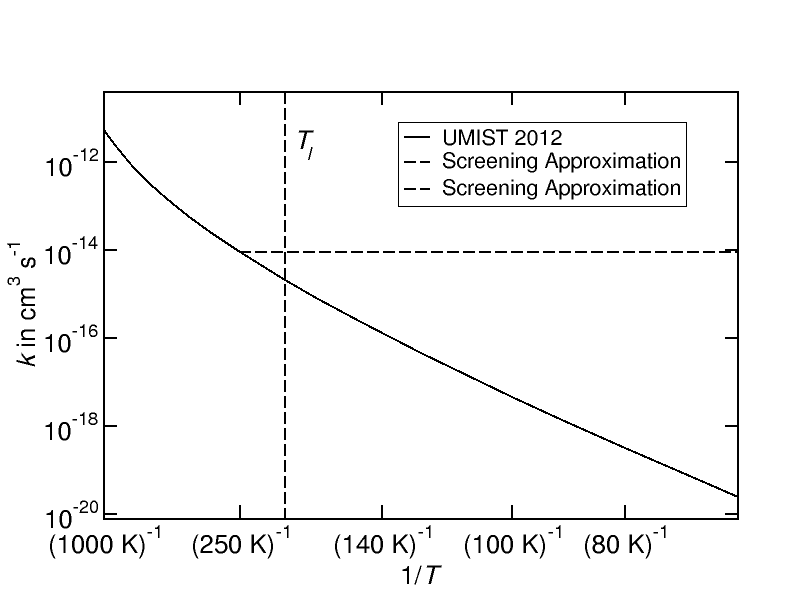}
\caption{Arrhenius plot for the reaction H$_2$~+~CN~$~\rightarrow$~HCN~+~H.
The solid line shows the UMIST2012 data; the dashed line represents the 
modification of the rate constants from 250~K on.
    The vertical dashed line indicates the lower temperature bound at which the measured or calculated data has been fitted: $T_l = 200$~K.
}
\label{fig:H2+CN-2}
\end{figure}

In order to estimate the impact of the effect, we use a
simple {ad hoc}
approach assuming that the original rate constant stays constant below a threshold temperature, thus breaking the original Arrhenius plot into two temperature regimes. 
Based on the experience of the reactions where tunneling corrected rate constants are available (reactions 1-3), we set the threshold temperature to $250$~K. For temperatures above 250~K, we keep the original UMIST2012 rate constant, and for the low-temperature part we use the rate constant of 250~K. 
This approach is presented in Fig.~\ref{fig:H2+CN-2} for one example reaction.
It is quite likely that this approach overestimates the tunnel effect; 
however, it is a suitable probe that can  determine for which reactions atom tunneling might be relevant.

For most chemical reactions, the rate constant still decreases at these temperatures and the 
rate constants do not become temperature independent until very low temperatures.
Therefore, this approximation can be assumed to overestimate the influence of atom tunneling
on the rate constants.
However, the approach is promising for the identification of reactions where tunneling could be important.

Table~\ref{tab:ScreeningRates} shows which rate constants were modified for the screening experiment.
For the screening, the standard disk around a young T Tauri star was used 
\citep{Woitke2016}.
After the screening,
a literature search was performed to check if the 
rate constants used in the UMIST2012 database already  
include atom tunneling appropriately. 
The chemical reactions identified via the screening where 
atom tunneling might be important are shown in 
Table~\ref{tab:CalculatedRates}.

\begin{table}
\caption{Reactions that could be accelerated by atom  tunneling according to our selection criteria. The reactions identified as interesting 
are  in bold. Chemical reactions are referred to as [No.].}
\vspace*{-2mm}
\begin{tabular}{lllllllll}
\hline\\[-3mm]
\hline
No. & \multicolumn{7}{c}{Reaction} \\
\hline\\[-3mm]
4 & C      &+&  NH         & $\rightarrow$  &   N      &+& CH  \\[1mm]
5 & CH$_2$  &  + &CH$_2$   &      $\rightarrow$  &   CH$_3$  &  + &  CH \\[1mm]
6 & CH$_2$  &  + &CH$_4$    &     $\rightarrow$  &   CH$_3$  &  + &  CH$_3$  \\[1mm]
7 & CH$_2$  &  + &CN        &  $\rightarrow$  &  HCN  &  + & CH  \\[1mm]
8 & CH$_2$  &  + &O         &  $\rightarrow$  &   OH   &  + & CH  \\[1mm]
9 & CH$_2$  &  + &OH        &  $\rightarrow$  &   H$_2$O  &  + & CH  \\[1mm]
10 & CH$_2$  &  + &OH        &  $\rightarrow$   &  O     & + &  CH$_3$ \\[1mm]
11 & CH$_3$  &  + &CH$_3$    &     $\rightarrow$  &   CH$_4$  &  + & CH$_2$ \\[1mm]
12 & CH$_3$  &  + &CN        &  $\rightarrow$  &  HCN  &   + & CH$_2$ \\[1mm]
13 & CH$_3$  &  + &H$_2$O    &    $\rightarrow$  &   OH  &    + & CH$_4$ \\[1mm]
14 & CH$_3$  &  + &NH$_3$     &   $\rightarrow$  &   CH$_4$   &  + & NH$_2$ \\[1mm]
15 & CH$_3$  &  + &OH       &  $\rightarrow$    &  CH$_4$  &   + & O   \\[1mm]
16 & CH$_3$  &  + &OH       &  $\rightarrow$   &   H$_2$O   & + & CH$_2$ \\[1mm]
17 & CH$_4$  &  + &CN       &  $\rightarrow$   &   HCN  &   + & CH$_3$ \\[1mm]
18 & CH     &+ &HCO       & $\rightarrow$   &   CO    &  + & CH$_2$  \\[1mm]
19 & CH     &+ &N         & $\rightarrow$   &   NH    &  +&  C  \\[1mm]
20 & CH     &+ &O        &  $\rightarrow$   &   OH    &  + &  C  \\[1mm]
21 & H$_2$   &  + &CH$_2$  &      $\rightarrow$ &     CH$_3$   &  + & H  \\[1mm]
22 & H$_2$   &  + &CH$_3$   &     $\rightarrow$ &     CH$_4$   &  + & H  \\[1mm]
23 & H$_2$   &  + &CH        & $\rightarrow$  &    CH$_2$   &  + & H \\[1mm]
\textbf{24} &\textbf{H}$_\textbf{2}$   &  +&  \textbf{CN}      &   $\rightarrow$  &   \textbf{HCN}    &+&  \textbf{H} \\[1mm]
25 & H$_2$     &+& NH      &   $\rightarrow$  &    NH$_2$    &+& H \\[1mm]
\textbf{26} & \textbf{H}$_\textbf{2}$    &+&  \textbf{O}       &   $\rightarrow$  &    \textbf{OH}     &+& \textbf{H}  \\[1mm]
27 & H      &+& CH$_3$     &   $\rightarrow$  &   CH$_2$    &+& H$_2$   \\[1mm]
\textbf{28} & \textbf{H}      &+& \textbf{CH}        & $\rightarrow$    &  \textbf{C}   &+& \textbf{H}$_{\textbf{2}}$  \\[1mm]  
29 & H      &+& H$_2$O     &   $\rightarrow$  &   OH     &+&  H$_2$   \\[1mm]
30 & H      &+& NH$_2$     &   $\rightarrow$  &    NH     &+&  H$_2$  \\[1mm]
31 & H      &+& NH       &  $\rightarrow$   &   N      &+&  H$_2$  \\[1mm]
32 & H      &+& OH       &  $\rightarrow$   &   O      &+&  H$_2$    \\[1mm]
33 & N      &+& HCO     &   $\rightarrow$   &   CO     &+&   NH  \\[1mm]
34 & NH$_3$  &+& H                      &  $\rightarrow$  & NH$_2$    &+& H$_2$ \\[1mm]
\textbf{35} &  \textbf{NH}   $_ \textbf{2}$    &+&  \textbf{H}$_ \textbf{2}$    &    $\rightarrow$  &      \textbf{NH}$_ \textbf{3}$   &+&   \textbf{OH}   \\[1mm]
36 & NH$_2$    &+& CH$_4$  &      $\rightarrow$  &    CH$_3$    &+&   NH$_3$   \\[1mm]
37 & NH$_2$    &+& OH       &  $\rightarrow$  &   NH$_3$    &+& O    \\[1mm]
38 & NH     &+& CH$_4$       & $\rightarrow$  &    CH$_3$    &+&  NH$_2$    \\[1mm]
39 & NH     &+& CN      &   $\rightarrow$   &   HCN    &+&  N     \\[1mm]
40 & NH     &+& OH       &  $\rightarrow$    &  NH$_2$    &+&   O   \\[1mm]
\textbf{41} &  \textbf{O}      &+&  \textbf{CH}$_ \textbf{4}$    &    $\rightarrow$  &     \textbf{OH}     &+&  \textbf{CH}$_ \textbf{3}$  \\[1mm]
42 & O      &+& H$_2$O     &   $\rightarrow$  &    OH     &+&  OH   \\[1mm]
43 & O      &+& NH$_3$      &  $\rightarrow$  &    OH     &+&  NH$_2$  \\[1mm]
44 & OH     &+& CN      &   $\rightarrow$  &    HCN    &+&   O    \\[1mm]
45 & OH     &+& HCN      &  $\rightarrow$  &   CN       &+&  H$_2$O  \\[1mm]
46 & OH     &+& NH$_3$    &    $\rightarrow$ &    H$_2$O   &+&  NH$_2$   \\[1mm]
47& OH     &+& OH       &  $\rightarrow$   &   H$_2$O    &+&   O  \\[1mm]
48 & H$_2$    &+& C$^+$         &  $\rightarrow$  & CH$^+$      &+& H \\[1mm]
49 & H$_2$    &+& N$^+$         &  $\rightarrow$  & NH$^+$    &+& H \\[1mm]
50 & H        &+& CH$_2^+$      &  $\rightarrow$  & CH$^+$    &+& H$_2$ \\[1mm]
51 & H        &+& CH$_3^+$      &  $\rightarrow$  & CH$_2^+$  &+& H$_2$ \\[1mm]
\hline
\end{tabular}
\label{tab:ScreeningRates}
\end{table}


\subsection{\jm{Planet-forming} disk models}
\label{sect:DiskModels}

We use the thermo-chemical disk modeling code ProDiMo \citep{Woitke2009, Kamp2010,Thi2011a, Aresu2012} to calculate the chemistry in steady state in a standard disk around a young T Tauri star and a young Herbig Ae star. The disks have very different geometries and central stars and thus provide two complementary environments to study the effects of atom tunneling in disks. The basic physical model structure is described for the T Tauri disk in \citet{Woitke2016} and for the Herbig disk in \citet{Tilling2012}. We repeat here only some of the key features of the two models.
The dust temperatures and the local radiation field are calculated from 2D radiative transfer using the standard DIANA dust opacities \citep{Min2016a}. We keep the underlying density and temperature structure of the reference models fixed when changing individual reaction rate constants. The computational grid is 90x90 (radial~versus~vertical points).

The chemical rate constants are taken primarily from the UMIST2012 database \citep{McElroy2013b}, but photorates are calculated using the local radiation field and cross sections from the Leiden database \citep{vanDishoeck2008} and reactions of excited H$_2$ \citep{Kamp2017}.  In the screening, we restrict ourselves to the small chemical network with 100 species and 1299 reactions described by \citet{Kamp2017}. 
In the subsequent study with updated rate constants from the literature, we use the large chemical network with 235 species and 3167 reactions. 
It is important to note that we generally extrapolate rate constants beyond the temperature range of validity except for lower temperatures in the case of negative $\gamma$ or higher temperatures for positive $\gamma$ (this would lead to a divergence of the rate constant).

Since we focus here entirely on \jm{gas-phase} chemistry, we exclude for the remainder of the work the ice reservoir of the disk. To identify the location of the ice reservoir we locate regions where at least one monolayer of ice has been deposited on the grain surfaces. These regions are subsequently excluded in the analysis of differences in species abundances. 

\begin{figure*}[htb]
\begin{center}
\includegraphics[width=8.cm]{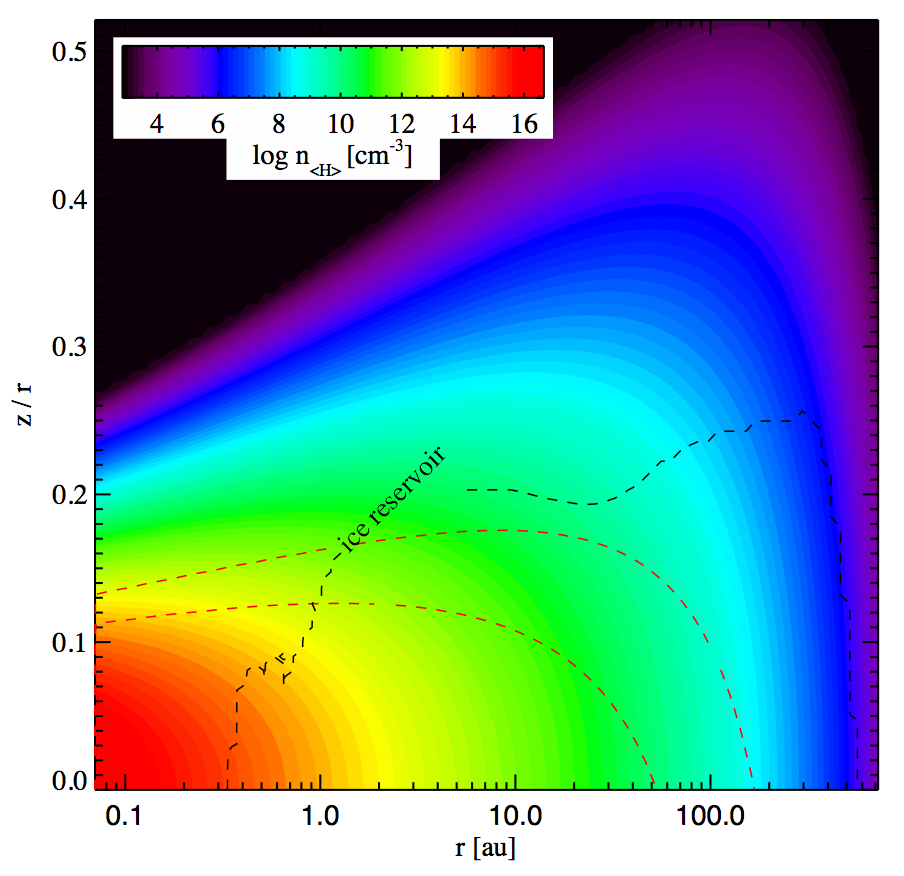}
\includegraphics[width=7.9cm]{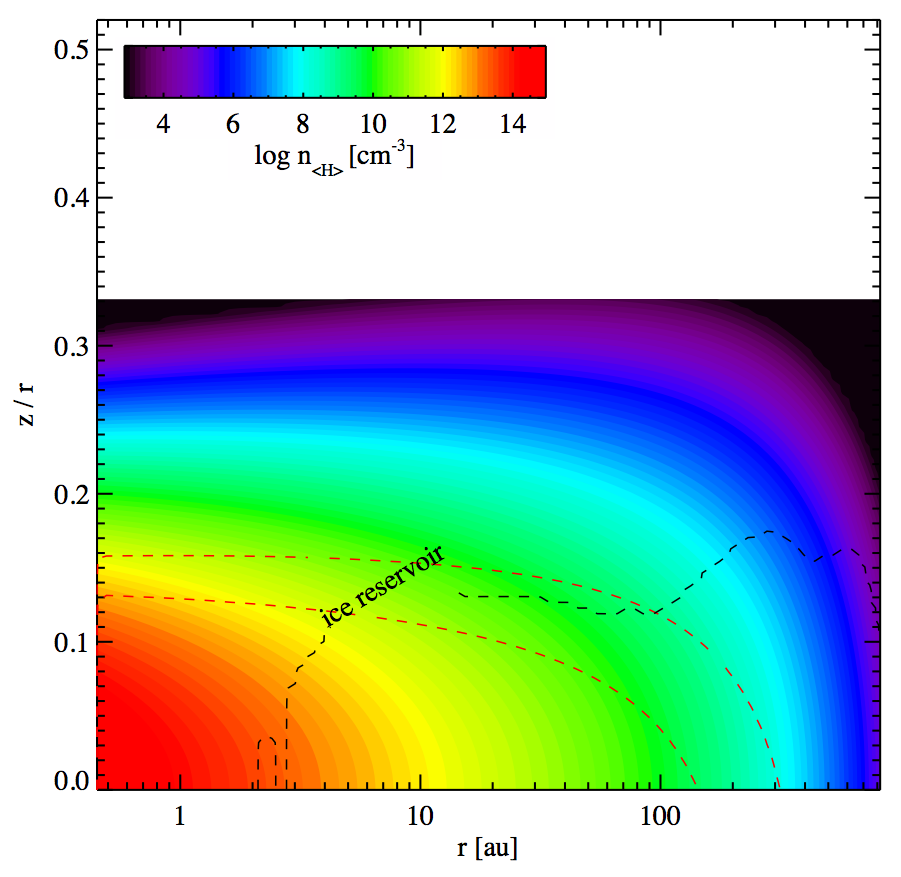}
\includegraphics[width=8.2cm]{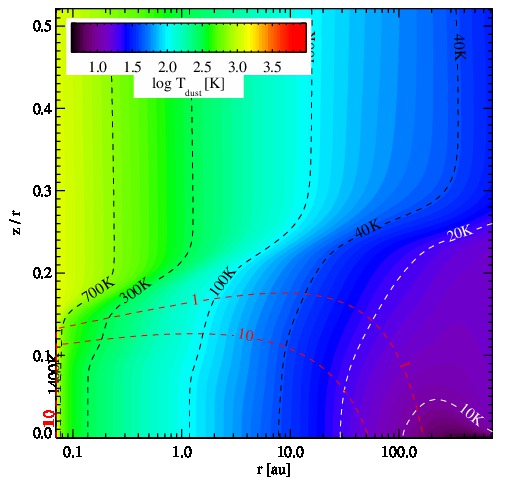}
\includegraphics[width=8.cm]{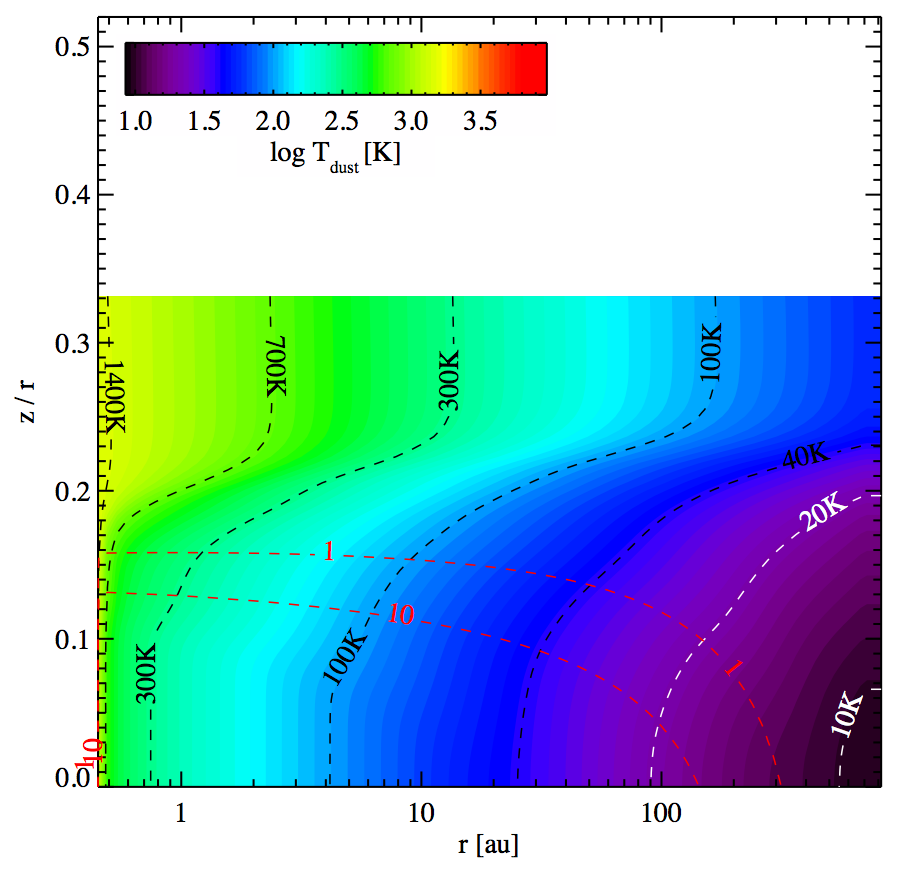}
\includegraphics[width=8.2cm]{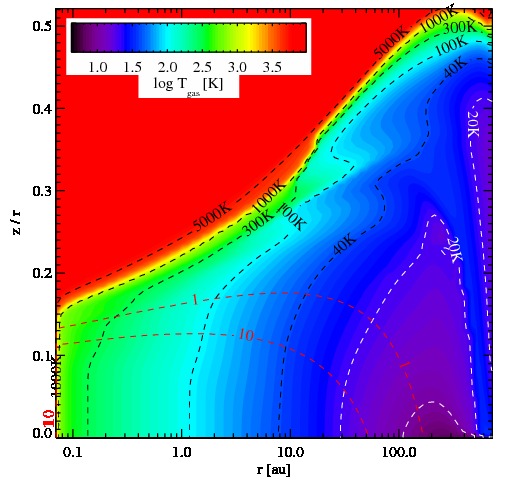}
\includegraphics[width=8.1cm]{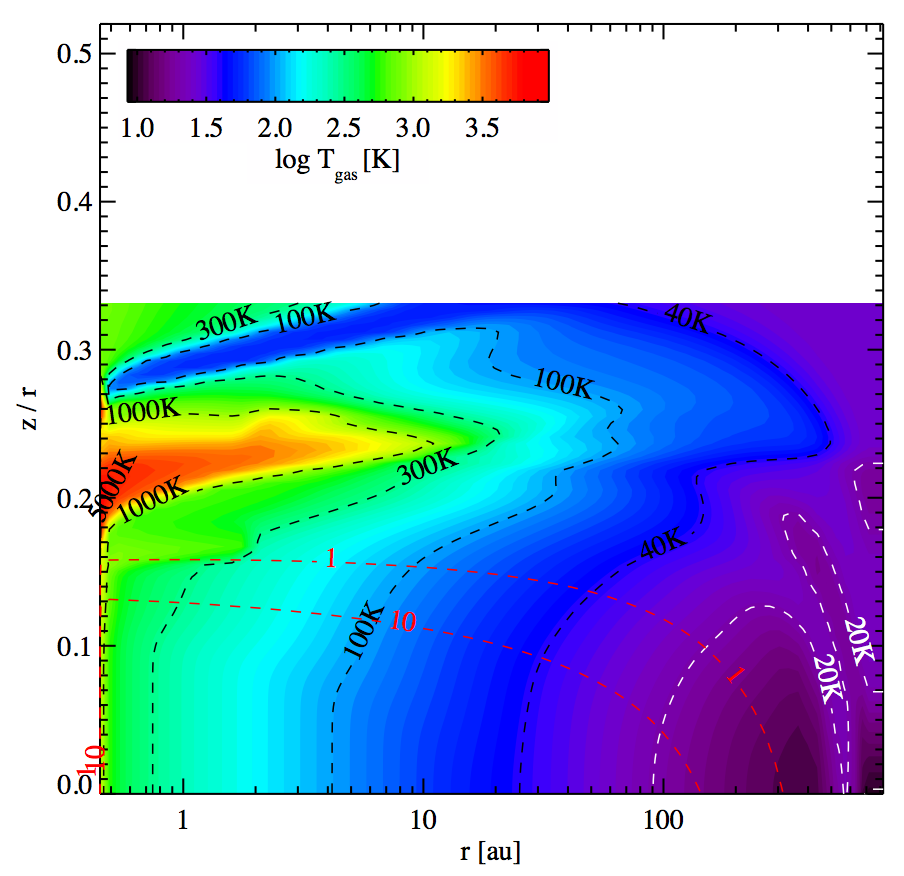}
\end{center}
\caption{Two-dimensional structure of the T Tauri (left) and Herbig (right) reference disk models. From top to bottom: Total hydrogen number density,  dust  temperature,  gas temperature. The black dashed contour on the density panels outlines the ice reservoir as defined by one monolayer of ice. The red dashed contours are the minimum of radial and vertical $A_{\rm V}=1$ and 10.}
\label{fig:diskdenstemp}
\end{figure*}

Figure~\ref{fig:diskdenstemp} shows the density and temperature (dust and gas) structure of the two reference models. The T Tauri disk model shows a flaring structure: the height of the $A_{\rm V}=1$ dashed contour increases with distance from star $r$. 
In the disk atmosphere above $A_{\rm V}=1$ 
gas and dust temperatures decouple; below $A_{\rm V}=10$ the disk model is vertically isothermal. The Herbig disk model on the other hand has a flat surface, i.e.,\ the disk is not strongly flaring. Above the $A_{\rm V}=1$ contour the temperature regime interesting for tunneling in the gas phase ($140~K<T_{\rm gas}<250$~K) is more radially extended than in the T Tauri disk model; at lower temperatures water freezes out and above
250~K tunneling is not important. 

%

\subsection{Identification of the importance of a chemical reaction\label{sub:criterium}}
We take a previously published reference model and then study the impact of individually changed reaction rate constants on the steady-state chemistry by comparing abundances of all species ($\epsilon_i=n_i/n_{\langle H\rangle}$, where $n_{\langle H\rangle}$ is the total hydrogen number density) to those of a reference model $\left(\epsilon_i^{\rm ref}\right):$
\begin{equation}
\Delta \log \epsilon_i = \log \left( \frac{\epsilon_i}{\epsilon_i^{\rm ref}} \right)\,\,\, .
\end{equation}
We evaluate the distribution function of abundance differences $\Delta \log \epsilon_i$ and flag a rate if the distribution is wider than the numerical noise. 
Since we restrict ourselves to the study of gas-phase reaction rates, we exclude the ice reservoir (as defined in Sect.~\ref{sect:DiskModels}) from  further analysis.

As a second step, we study the impact of those changes on the emission line fluxes using a set of standard lines of various ionic/atomic/molecular species (e.g., atomic fine structure lines, rotational lines of H$_2$, CO, OH, H$_2$O, HCO$^+$, N$_2$H$^+$) across a wide wavelength range (optical to submm). We use the same 56 standard lines as \citet{Kamp2017}. If a reaction is found to have significant changes on species abundances, we also investigate the resulting changes in line fluxes with respect to the standard model.

\subsection{Arrhenius fits}
%
%

To obtain a more reliable and meaningful result,
we fitted the parameters of 
the modified Arrhenius equation (Eq.~\ref{eq:modArrh}) for 
 reactions [1], [2], and [3]
and  for the reactions where the screening 
showed a significant change in species abundances
(see the above-mentioned criteria).
Again, a two-segment Arrhenius plot is constructed, 
divided by a reaction-dependent threshold temperature $T_{\rm thresh}$.
Below $T_{\rm thresh}$ the Arrhenius parameters are fitted to 
the literature values obtained by quantum chemistry. 
The fits were performed 
using  the  nonlinear least-squares (NLLS) Marquardt--Levenberg algorithm
implemented in gnuplot~v~5.0 \citep{gnuplot}. 
For the  high-temperature regime, i.e., for $T > T_{\rm thresh}$,
the UMIST2012 values were used.

As the instanton rate constants close to 
the crossover temperature $T_{\text{C}}$ are overestimated, 
we use the values below  
$T_{\rm thresh} \approx 0.9~T_{\text{C}}$ for the fitting of the modified Arrhenius parameters for the reactions [1], [2], and [3].
For the other reactions, a literature survey for the corresponding low-temperature rate constants is performed.
For these reactions, the threshold temperatures  $T_{\rm thresh}$ 
listed in Table~\ref{tab:CalculatedRates} were  used. 
If the two rates do not match at the intersection $T_{\rm thresh}$, ProDiMo performs a weighted average to ensure a smooth rate constant.



\clearpage

\section{Results}

To investigate the influence of the tunneling effect, 
we first present results from a pilot study on the reaction H$_2$~+~OH~$\rightarrow$~H$_2$O~+~H~[1] comparing the species densities in a T~Tauri disk model using the rate constants calculated without tunneling to those using the rate constants from the UMIST2012 database. We next present which of the 47 reactions that are susceptible to tunneling (Table~\ref{tab:ScreeningRates}) cause major deviations in the species densities in the T Tauri disk model. For these reactions, we present here rates collected from the literature. In the last step, we perform a detailed comparison between the species densities from disk models calculated with the revised literature rates and instanton theory (Table \ref{tab:CalculatedRates}) and those calculated with UMIST2012 rates.

\subsection{Influence of tunneling in the reaction H$_2$~+~OH~$\rightarrow$~H$_2$O~+~H on disk models}

We use the rate constants for this reaction calculated in the absence of tunneling (Sect.~\ref{sect:examplewaterreaction}) and compare the results for a T~Tauri disk model to those of the same disk model using the original UMIST2012 values of the rate constants.

\begin{figure*}[thb]
\includegraphics[width=9cm]{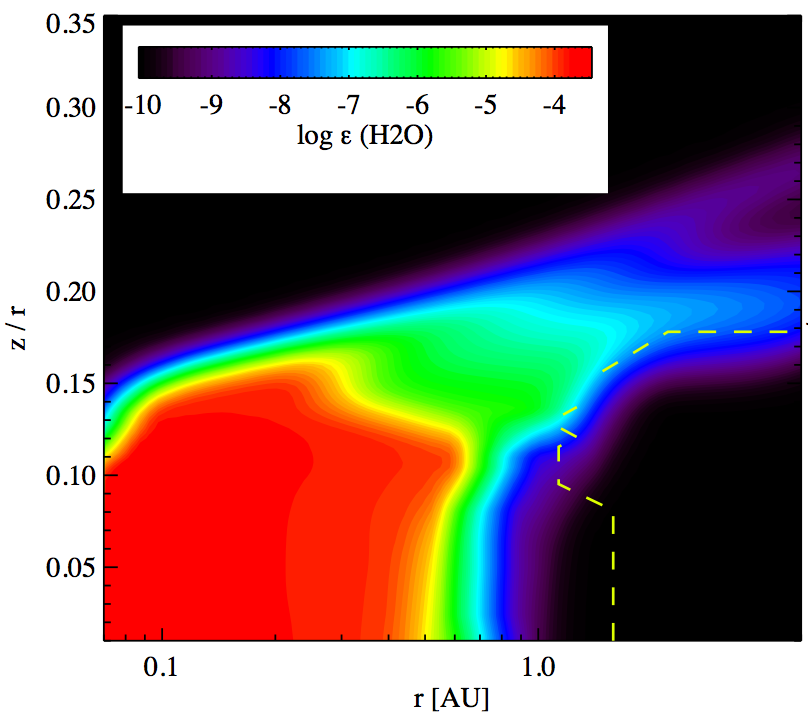}
\includegraphics[width=9cm]{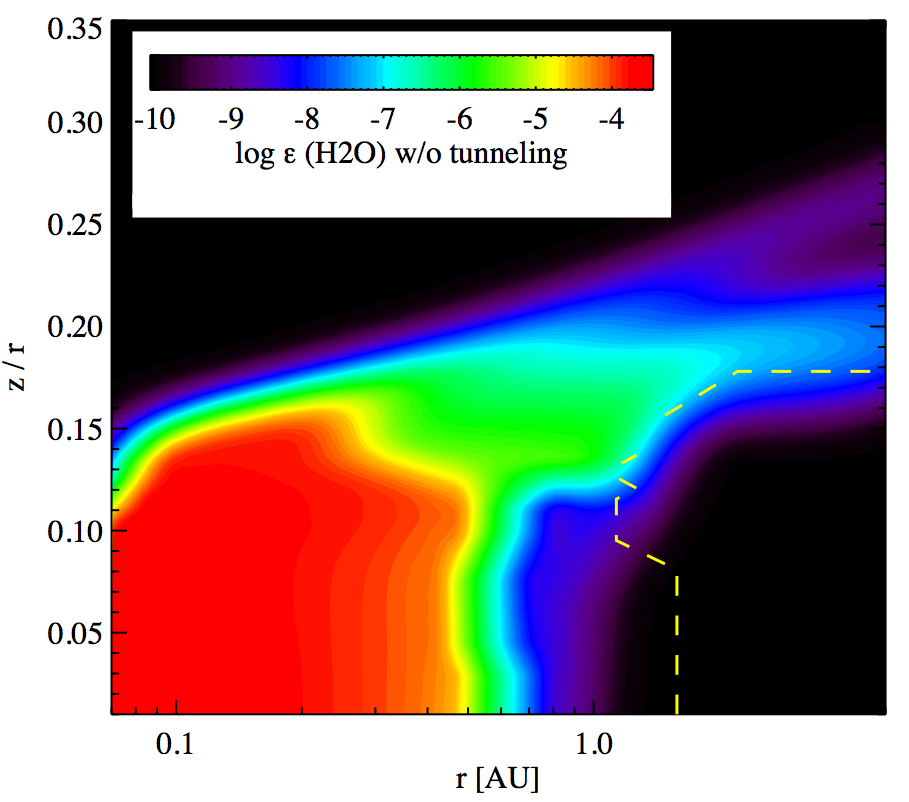}
\caption{Left: Water abundance in the T Tauri disk model using the rate constant for H$_2$~+~OH~$\rightarrow$~H$_2$O~+~H from UMIST2012. Right: Same, but now using a rate constant calculated without tunneling. The yellow dashed contour outlines the region below which 300 monolayers of ice exist.}
\label{fig:TTAURI_H2O_tunnelnotunnel}
\end{figure*}

\begin{figure}[thb]
\includegraphics[width=9cm]{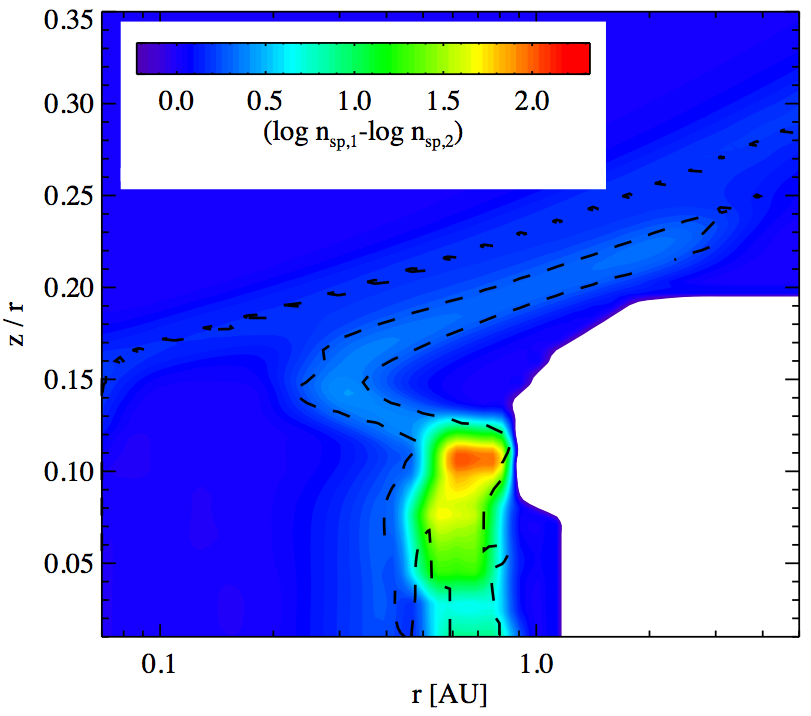}
\caption{Difference in H$_2$O species density between the model using the UMIST2012 rate constants and the model using the rate excluding the effect of tunneling for reaction [1]: H$_2$~+~OH$~\rightarrow~$H$_2$O~+~H. The black dashed line shows a difference of 0.5~dex.}
\label{fig:TTAURI_H2O_diff}
\end{figure}

Figure~\ref{fig:TTAURI_H2O_tunnelnotunnel} shows a zoomed image of  the T~Tauri disk model around the snowline at $\sim 1$~au. Water species density values can differ by  up to a factor of $\sim 200$ (Fig.~\ref{fig:TTAURI_H2O_diff}) close to the snowline at $z/r\sim 0.1$. However, even in surface layers, there can be deviations of up to a factor of 3. In the disk surface, where photons provide a certain level of ionization, the main pathways of water formation are through ion-molecule chemistry or radiative association H~+~OH$~\rightarrow$~H$_2$O \citep[e.g.,][]{Woitke2009b,Najita2011,Kamp2013}. None of those reaction pathways shows significant tunneling effects, and hence we expect most of the disk water reservoirs to remain unchanged. 

\begin{figure}[thb]
\includegraphics[width=9cm]{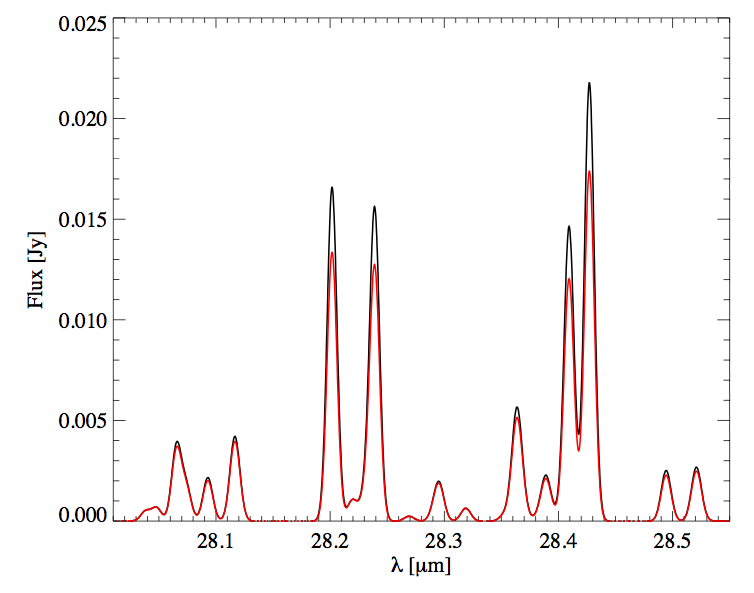}
\caption{Water spectrum from the model with UMIST2012 rate (black) and without tunneling in the rate constant of the reaction
H$_2$~+~OH$~\rightarrow$~H$_2$O~+~H (reaction [1], red).
}
\label{fig:mid-IRwaterspectra}
\end{figure}

This inner disk water reservoir around the position of the radial midplane snowline gives rise to mid-IR spectra. Observations with the Spitzer Space Telescope have shown that many disks are rich in rotational and ro-vibrational water lines \citep[e.g.,][]{Carr2008,Pontoppidan2010}. Figure~\ref{fig:mid-IRwaterspectra} shows how these changes in water abundance propagate into the mid-IR spectrum. Line fluxes can deviate by up to 60\% (Fig.~\ref{fig:compare-lines-ref-notunnel}), but more importantly, water lines are often very optically thick and originate at very different radii and different depth, thus changing also the relative fluxes and the overall appearance of the spectrum. 
Changes affecting only  strong lines, for example,  can influence the temperature determination from such a spectrum using slab models. 

In the following, we perform a detailed screening of all reactions in the small chemical network to identify rates susceptible to similar tunneling effects.

\clearpage

\subsection{Screening of chemical reactions}

Using the criteria defined in Sect~\ref{sub:criterium}, we identified 47 
reactions that could show tunneling effects. As stated above in the methods section, this 
screening strongly overestimates the influence of atom tunneling on the chemical reactions under consideration. We compare the results from the T Tauri disk model for each reaction individually. This leads to four additional reactions that show differences in species densities and that could be important for the \jm{gas-phase} chemistry in disks (marked bold in Table~\ref{tab:ScreeningRates})
\begin{align}
{\rm H_2 + CN} & \rightarrow {\rm HCN + H} & [24] \nonumber \\
{\rm H_2 + O} & \rightarrow {\rm OH + H} & [26] \nonumber \\
{\rm H + CH} & \rightarrow {\rm C + H_2} & [28] \nonumber \\
{\rm NH_2 + H_2} & \rightarrow {\rm NH_3 + H} &[35] \nonumber \\
{\rm O + CH_4} & \rightarrow {\rm OH + CH_3} &[41] \nonumber 
\end{align}

\begin{figure*}[thb]
\includegraphics[width=9cm]{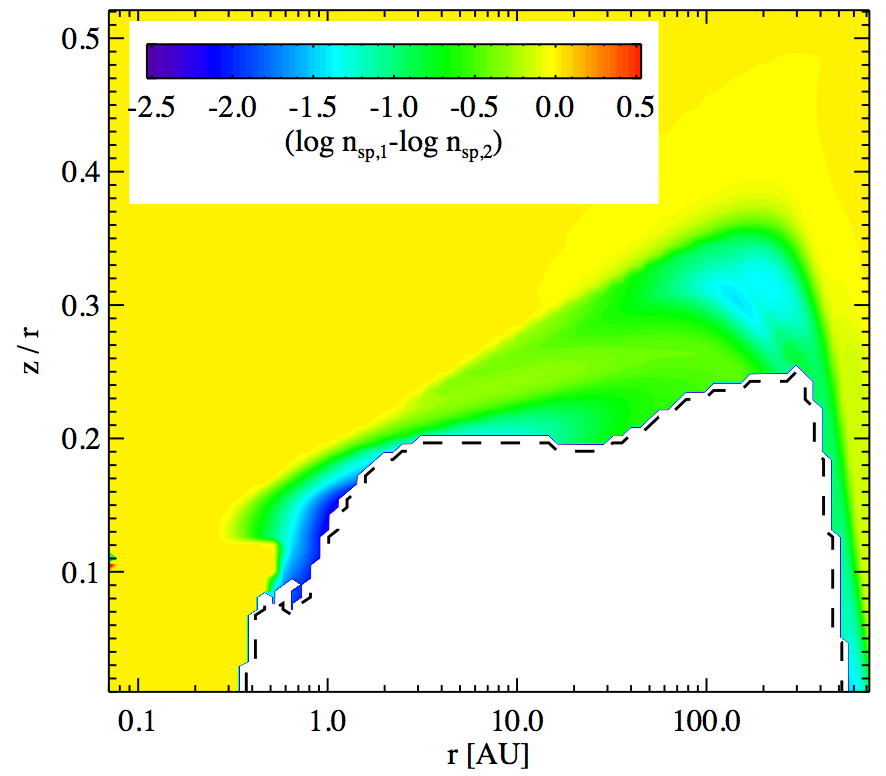}
\includegraphics[width=9cm]{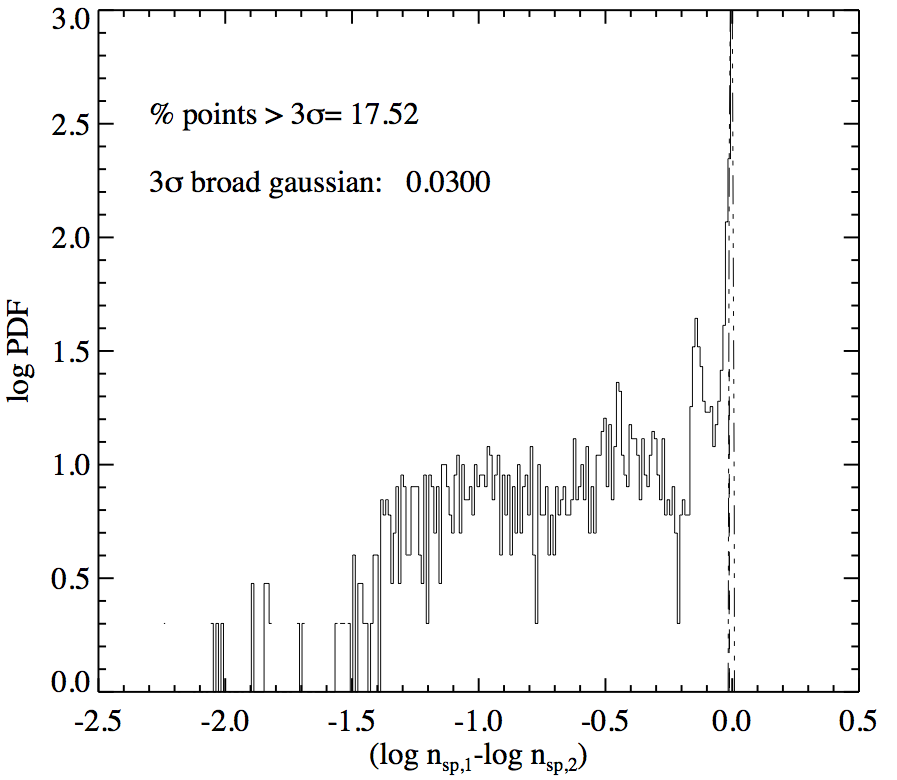}
\caption{Screening test. Left: Difference in HCN species density between the reference model ($n_{\rm sp,1}$) and the model with modified reaction 
H$_2$~+~CN~$\rightarrow$~HCN~+~H
($n_{\rm sp,2}$, reaction [24]). The black dashed line indicates the surface of the ice reservoir excluded from the statistical analysis. Right: Distribution function of abundance differences with Gaussian fit to the numerical noise (dash-triple-dotted line).}
\label{fig:HCN-diffs}
\end{figure*}

We use reaction [24] to discuss the influence of a 
temperature-independent rate constant below 250~K on
the results of the disk chemistry simulations. We monitor the difference in HCN abundance in the T~Tauri disk model and the distribution function for $\Delta \log \epsilon_{\rm HCN}$ (see Fig.~\ref{fig:HCN-diffs}).
Roughly 18\% of the grid points outside the ice reservoir (black dashed line) show a deviation of  more than 3$\sigma$  from the numerical noise. The changes occur predominantly in the upper disk layers beyond 20~au. However, also at the outer edge of the disk, the HCN density increases due to the enhanced rate constant up to a factor of $\sim 30$. 

\subsection{Use of correct reaction rate  constants}

\begin{table*}[htb]
\caption{Reactions with recently calculated low-temperature rate constants, their references
(including the lower boundary of the recommended temperature range of the UMIST2012 database, $T_l$)}, their threshold temperature $T_{\rm thresh}$, and their respective Arrhenius fit parameters for the low-temperature regime below $T_{\rm thresh}$.
\begin{tabular}{lll|l|lll|l|lll}
\hline\\[-3mm]
No. & Reaction & Ref. & $T_{\rm thresh}$ & \multicolumn{3}{c|}{UMIST2012 Values}& $T_l$ & \multicolumn{3}{c}{low-temperature fit}\\
       &          &           & [K] & $\alpha$ & $\beta$ & $\gamma$ & [K] &$\alpha$ & $\beta$ & $\gamma$ \\
\hline
\hline\\[-3mm]
1 & H$_2$~+~OH~$\rightarrow$~H$_2$O~+~H & M16 & 200 & 2.05(-12) & 1.52  & 1736  &250 & 2.342(-15) & 5.2729 & 0.072 \\[1mm]
2 & CH$_4$~+~OH~$\rightarrow$~CH$_3$~+~H$_2$O & L17 & 240 & 3.77(-13) & 2.42 & 1162 &178 & 8.252(-14) & 7.753 & 0.040\\[1mm]
3 & CH$_4$~+~H~$\rightarrow$~CH$_3$~+~H$_2$ & B16 & 300 & 5.94(-13) & 3.0 & 4045 &300 & 1.19(-19) & 16.37 & 1.217\\[1mm]
\hline\\[-3mm]
\multicolumn{9}{c}{literature search after screening}\\
\hline
\hline\\[-3mm]
24 & H$_2$~+~CN~$\rightarrow$~HCN~+~H & J06  & 200 & 4.04(-13) & 2.87 & 820 &200 & 1.826(-13) & 11.920 &  6.3(-4) \\[1mm]
26 & H$_2$~+~O~$\rightarrow$~H~+~OH & B03 & 450 & 3.14(-13)& 2.70 & 3150 &298 & 5.79(-16) & 7.077 & 1340  \\[1mm]
41 & CH$_4$~+~O~$\rightarrow$~OH~+~CH$_3$ & Z16 & 500  & 2.29(-12)  &2.20 & 3820 &298 & 7.132(-15) &5.200  & 1560  \\[1mm]
\hline
\end{tabular}
\label{tab:CalculatedRates}
\tablefoot{
The notation $x(-y)$ stands for $x \cdot 10^{-y}$. References: 
M16  \citep{meisner2016}, 
L17 \citep{lamberts2017}, 
B16 \citep{beyer2016},
J06 \citep{ju2006},
B03 \citep{balakrishnan2003},
Z16 \citep{zhao2016}.
}
\end{table*}

We combed through the literature to obtain more realistic rate constants for these five reactions. 
The UMIST2012 data for reaction [28] is already in  excellent agreement with the recent 
quantum dynamical results of \citet{gamallo2012}. Thus, we do not include this reaction in the subsequent work.
For reaction [35], the UMIST2012 values overestimate the reaction rate compared to the computations by \citet{nguyen2019},
as can be seen in figure~\ref{fig:NH2+H2}.
The recently published values based on highly accurate 
semiclassical transition state theory (SCTST) calculations are state of the art and are shown to deviate just slightly from experimental values.
Therefore, this reaction cannot be used to elucidate whether atom tunneling  plays
a crucial role and will be omitted. 
It should  be noted, however, that the reaction rate constants computed on CVT/$\mu$OMT-Level might underestimate atom tunneling in the lower temperature range.  We therefore  encourage  updating the Arrhenius parameters of this reaction in the next version of the UMIST database

In the following, we briefly discuss the influence of atom tunneling on the rate constants of  reactions [1], [2], and [3], and on  those identified in the screening, reactions [24], [26], and [41].
For these six reactions, we computed Arrhenius fits following the method outlined in Sect.~2.4. The resulting parameters are summarized in Table~2 (bottom).
We then  present results when using Arrhenius parameters fitted to these new rate constants.
The analysis method is the same as above.
We want to highlight the vanishingly small values of
the fitting parameter $\gamma$ 
for the low-temperature fit of  exothermic reactions [1], [2], [3], and [24], which have little impact on the rate constant.


\subsubsection{H$_2$~+~OH~$\rightarrow$~H$_2$O~+~H}

\begin{figure*}[htb]
\includegraphics[width=9cm]{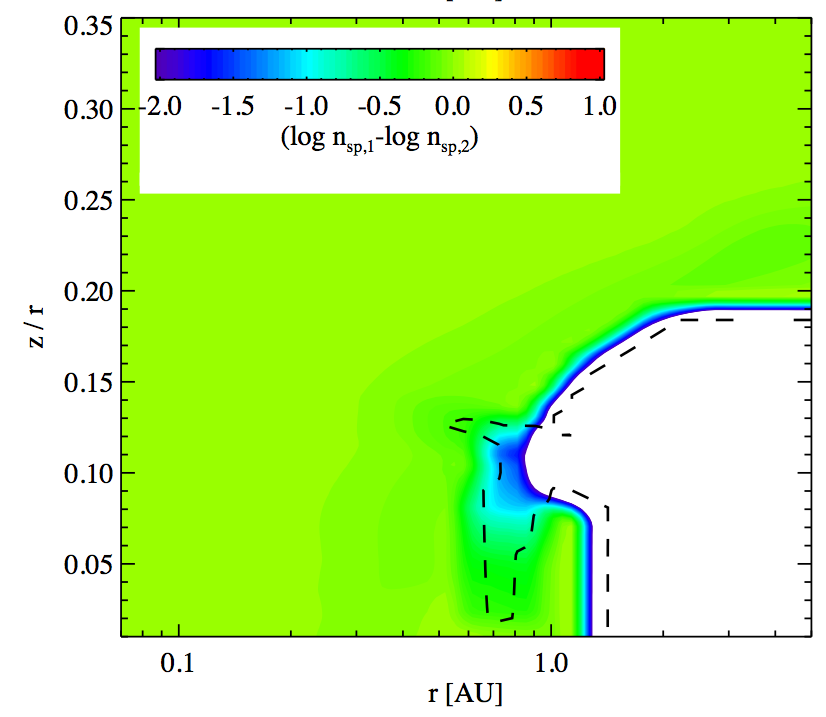}
\includegraphics[width=9cm]{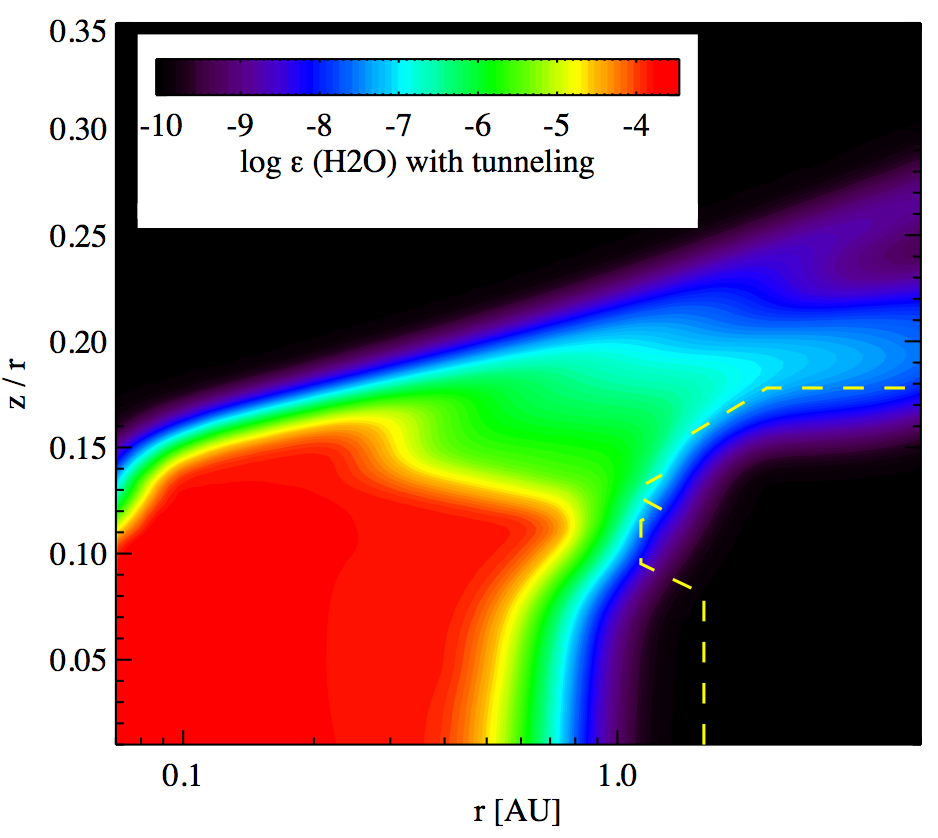}
\caption{Difference in water density (left) between the reference T Tauri model ($n_{\rm sp,1}$) and the model with new rate constants with accurate description of atom tunneling ($n_{\rm sp,2}$) for the reaction 
H$_2$~+~OH~$\rightarrow$~H$_2$O~+~H according to Table~\ref{tab:CalculatedRates}. The black dashed line shows a difference of 0.5~dex in water abundance in the T Tauri disk model that uses the new rate constants (right); this is to be compared to Fig.~\ref{fig:TTAURI_H2O_tunnelnotunnel} (we note the different scale). The yellow dashed contour outlines the region below which 300 monolayers of ice exist. }
\label{fig:TT-H2O-diffs-calculated}
\end{figure*}

As previous work in \citet{meisner2016} shows, atom tunneling 
increases the rate constant for this water forming reaction by several orders of magnitude (see
Fig.~\ref{fig:H2+OH}).
However,  using the new Arrhenius parameters, the water abundance (density) increases just above the ice reservoir ($z/r<0.3$) in the T Tauri disk at distances of 0.3--50~au from the central star (see Fig.~\ref{fig:TT-H2O-diffs-calculated}).
The increase in water abundance is typically less than a factor two. The spatial extent of this region is largely limited by ion-molecule chemistry dominating the water formation closer to the disk surface (higher $z/r$) where photons can reach. In addition, the rate constant increase is noteworthy only below $\sim 200$~K, which also limits spatially the impact of the enhanced water formation. In addition, surface chemistry, especially  chemical desorption from surfaces, can enhance the gas-phase water abundance around the snowline \citep{Cazaux2016}; however, this process needs to be studied in a future work.

\begin{figure*}[htb]
\includegraphics[width=9cm]{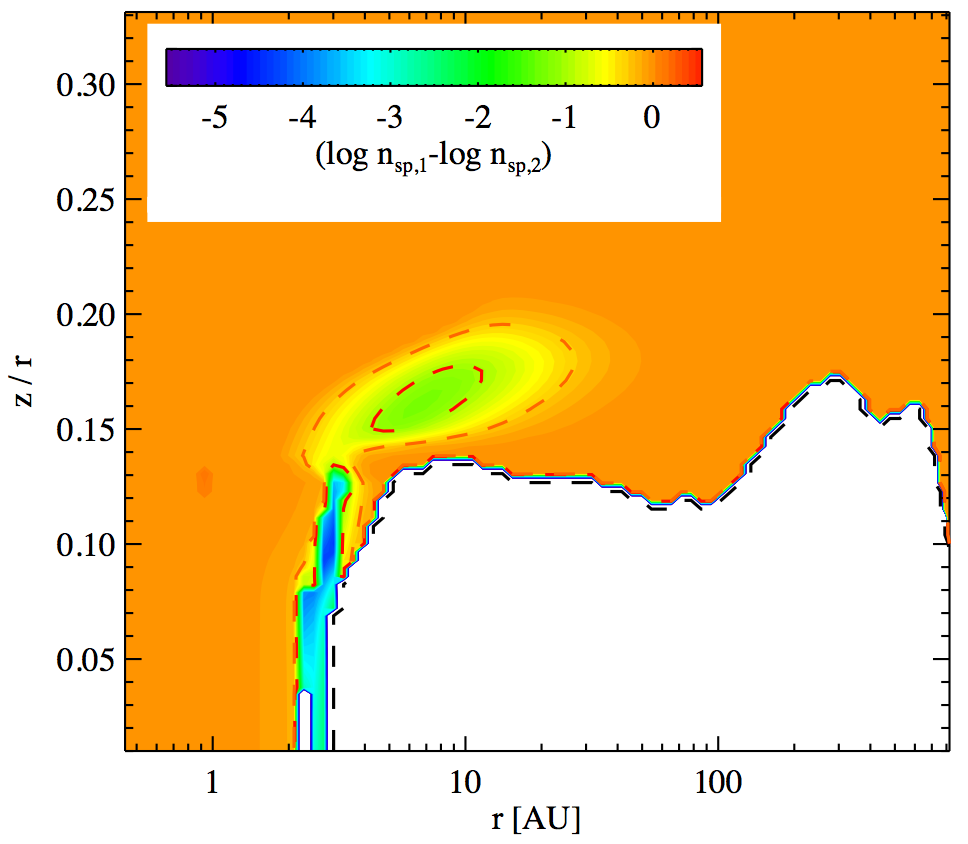}
\includegraphics[width=9cm]{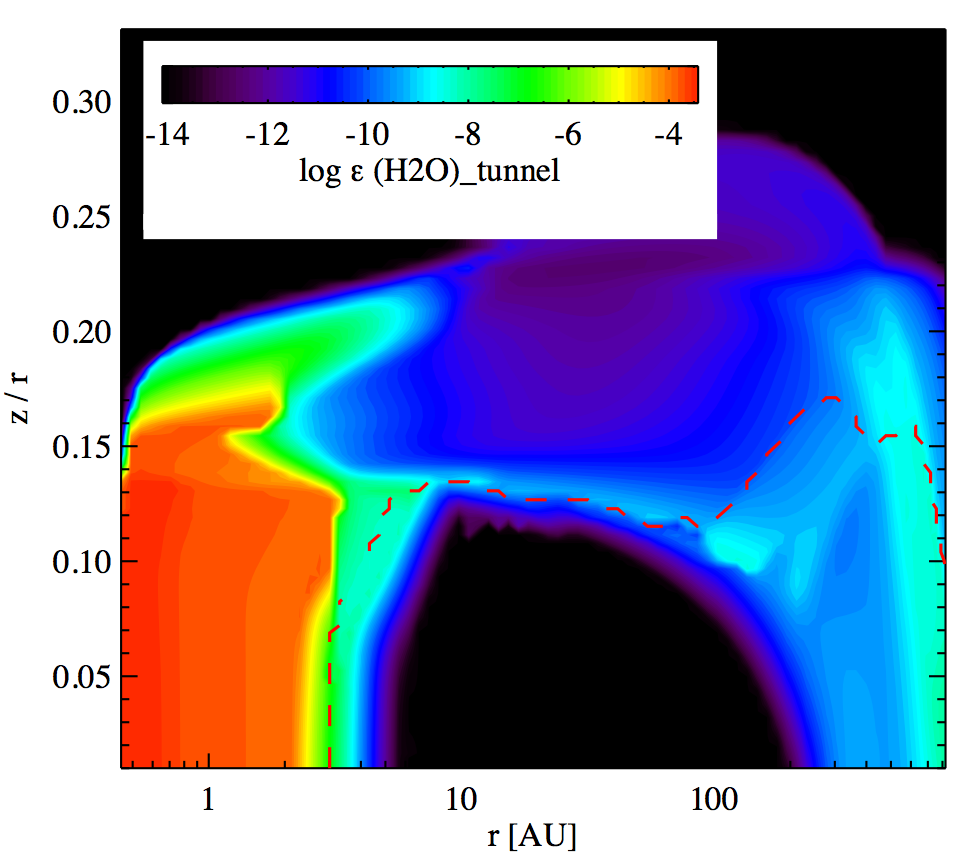}
\caption{Difference in water density between the reference Herbig model ($n_{\rm sp,1}$) and the model with new rate constants with accurate description of atom tunneling ($n_{\rm sp,2}$) for the reaction 
H$_2$~+~OH~$\rightarrow$~H$_2$O~+~H according to Table~\ref{tab:CalculatedRates}. The black dashed line indicates the surface of the ice reservoir excluded from the statistical analysis, the red dashed line a change in species density of a factor of three. Water abundance (right) in the Herbig disk model that uses the new rate constants; the red dashed line indicates the surface of the ice reservoir.}
\label{fig:Herbig-H2O-diffs-calculated}
\end{figure*}

The Herbig disk model is flatter than the T Tauri disk model and has a more extended region where the temperature ranges between 200~K and the freeze-out of water. It is in this relatively cold region around 10~au that the water abundance increases strongly (more than a factor of 2-10) (see Fig.~\ref{fig:Herbig-H2O-diffs-calculated}). However, in these regions, the water abundance is very low ($\ll 10^{-8}$) and the higher rate constants do not improve the efficiency of water formation above abundances of $10^{-8 ... -9}$ (typical values reached by photodesorption in the outer disk).

\subsubsection{CH$_4$~+~OH/O/H~$\rightarrow$~CH$_3$~+~H$_2$O/OH/H$_2$}

\citet{zhao2016} calculated rate constants for the  reaction O~+~CH$_4 \rightarrow$~OH~+~CH$_3$ [41] using the quantum instanton method.
For the reaction of CH$_4$ with OH and H we used 
results based on semi-classical instanton theory  
from \citet{lamberts2017} and \citet{beyer2016}, respectively.
When using the fits to these rate constants in the T~Tauri disk model, 
we did not find significant differences in the species abundances. If differences occurred, they were restricted to a narrow radial range inside the inner edge of the ice reservoir. For most molecules that are affected, the abundances in these regions are very small $<\,10^{-10}$.

\subsubsection{H$_2$~+~CN~$\rightarrow$~HCN~+~H}

Reaction [24] has already been discussed  in the screening approach. 
For this reaction, \citet{ju2006} 
published rate constants down to 100~K using variational transition state theory in combination with the small-curvature tunneling scheme.
Even though the authors of this computational study state that
``{the tunneling effects are [...] non-negligible over 100–200K},'' the reaction rate constants provided by \citet{ju2006} 
are even slightly smaller than the UMIST2012 values in this temperature regime (see Fig.~\ref{fig:H2+CN}). We note that in addition to  using  the calculated tunneling rate, here we also use  the large chemical network; this also affects the outline of the ice reservoir. This is important to provide reliable abundances for CN and HCN as discussed in \citet{Greenwood2017} and \citet{Kamp2017}\footnote{The use of the small network in the screening is a valid approach since we are only interested in differential changes modifying one rate at a time. However, when also calculating  emergent spectra and/or line fluxes, we do need to use the large network.}.

\begin{figure}[thb]
\includegraphics[width=9cm]{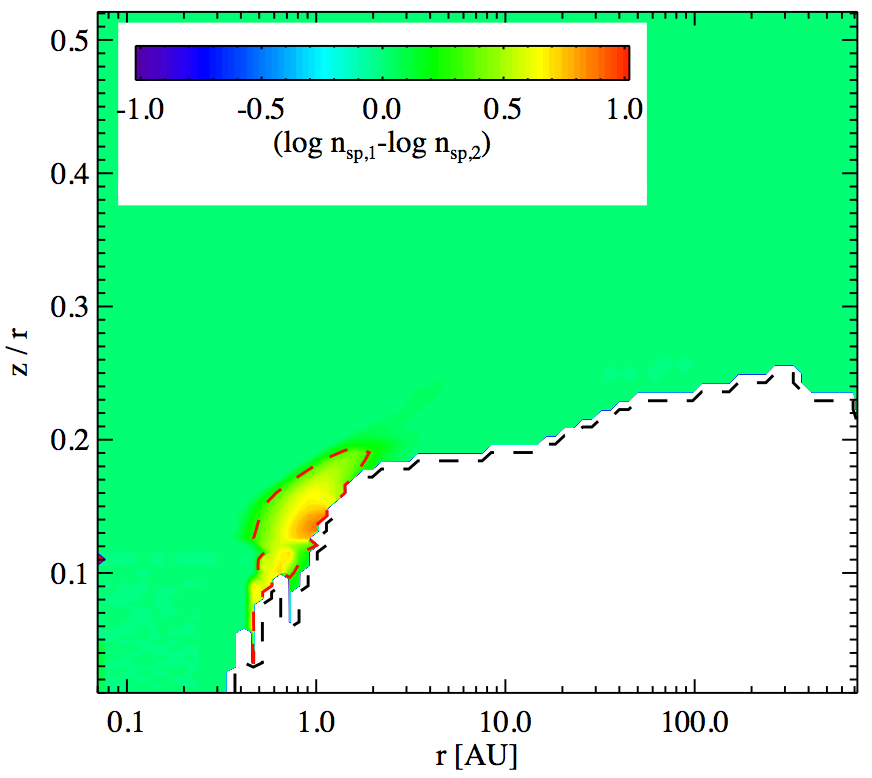}
\caption{
Difference in HCN density between the reference model ($n_{\rm sp,1}$) and 
the recent literature values of 
\citet{ju2006} for the reaction H$_2$~+~CN~$\rightarrow$~HCN~+~H ($n_{\rm sp,2}$) according to Table~\ref{tab:CalculatedRates} (reaction [24]). The black dashed line indicates the surface of the ice reservoir (one monolayer) excluded from the statistical analysis. The dashed red line shows the $\Delta \log \epsilon = 0.3$ contour. 
}
\label{fig:HCN-diffs-calculated}
\end{figure}

As a consequence of the lower rate constants compared to UMIST2012 values, the differences in HCN density become much smaller compared to the screening approach above and they also become  negative (Fig.~\ref{fig:HCN-diffs-calculated}). 
Most density changes stay well below a factor of three. HCN in the outer disk beyond 100~au does not change at all. While the reaction H$_2$+CN is the main HCN formation path in the upper surface layers of the disk, HCN is predominantly formed through C+NH$_2$ and N+CH$_3$ at the outer cold disk edge \citep{Greenwood2017}.
As a consequence of the small density changes, HCN line fluxes from low rotational lines do not change due to the new values of the rate constants for this reaction.

\subsubsection{H$_2$~+~O~$\rightarrow$~H~+~OH}

This reaction is endothermic, and therefore the
activation energy is not expected to  
decrease significantly at low temperatures
because endothermic reactions require energy to take place, 
leading to a nonvanishing activation energy
\citep{meisnerthesis2018}.
Nevertheless, in general, tunneling effects can also enhance reactivity for endothermic reactions.
However, \citet{balakrishnan2003} provides accurate 
rate constants for reaction [26] down to 200~K
that are in good agreement with the  UMIST2012 data, but  higher by a factor of  $\approx 4$ at 200~K.
It should  be noted that  the validity range provided in UMIST2012 just extends down to 297~K.
We therefore decided to  include this reaction in the study
even though tunneling is not as pronounced as it would be in an exothermic reaction.

Inclusion of atom tunneling in this reaction 
enhances the destruction of H$_2$ and leads to higher abundances of atomic hydrogen. It is obvious that this can transmit subsequently into the abundances of many other species including  hydrocarbons, for example. In most cases, the abundances of these species is small at the inner snowline; however, we find that CH$_4$, C$_2$H$_2$, C$_3$, and C$_3$H$_2$ as well as HCN, HNC, H$_2$CO, and CH$_3$OH can be abundant in this region and can be affected by this change in rate constant. Given the particular geometry/structure of our T Tauri disk model, this zone is relatively small radially and confined to $z/r<0.15$.

\subsection{Time-dependent chemistry versus steady state}

We  explored the neutral-neutral reaction 
H$_2$~+~OH~$\rightarrow$~H$_2$O~+~H also using time-dependent chemistry in the model with the small chemical network. In that case, we started with initial abundances taken from a molecular cloud run ($n_{\rm \langle H \rangle}= 10^4$~cm$^{-3}$, $A_{\rm V}=10$, and an age of $1.7\cdot 10^5$~yr). We then evolved the chemistry over an age of $3\cdot 10^7$~yr,  for the reference T Tauri disk model and for the model with the tunneling rate. The timescale for reaching steady state increases to a few Myr around the position of the snowline in the midplane. Above the ice reservoir in the disk surface, timescales are even shorter than $10^4$~yr. At timesteps of 0.3, 1, and 3 Myr, differences in the water abundance between UMIST2012 and the new instanton rate inside the snowline and below $z/r\sim0.1$ are slightly larger than in steady state, up to a factor two. 


 

\section{Astrophysical Implications}

\subsection{Emission lines from disks}

We compared the emerging line fluxes of the 56 rotational lines of H$_2$, CO, OH, H$_2$O, HCO$^+$, and N$_2$H$^+$ described in Sect.~\ref{sub:criterium} from the model with the increased rate constant for the reaction H$_2$~+~OH~$\rightarrow$~H$_2$O~+~H with the reference model. In both cases, T Tauri and Herbig disk, none of the line fluxes changes \jm{by more than a small percentage} which is well within the numerical accuracy of predicting them.
When using the updated Arrhenius parameters of the reactions
H$_2$~+~CN~$\rightarrow$~HCN~+~H [24]
and
H$_2$~+~O~$\rightarrow$~OH~+~H [26]
line fluxes also do not change beyond the numerical accuracy.

The line emission generally emerges from surface layers where the chemistry is dominated by ion-molecule reactions. Several molecular rotational lines can probe deeper layers in the outer disk; UV radiation can penetrate deeper in the outer disk, and hence  ion-molecule routes often also dominate here.
The relative small and spatially confined changes found above occur in regions that do not contribute to the series of lines selected here. Especially in the inner disk, many of the changes in species densities lie in the optically thick part of the disk, so well below the continuum optical depth of one.

\subsection{Planet formation}
\label{sub:planetform}

The \jm{C-to-O ratio} has been identified as a key quantity for linking disk models and exoplanets \citep[e.g.,][]{Madhusudhan2017}. It is a quantity that can be inferred from observed exoplanet atmosphere spectra \citep[see a recent review by][]{Deming2017} and is often found to be consistent with solar \jm{C-to-O ratio} \citep[e.g.,][]{Line2014}. Also, it is readily extracted from simple or more complex disk models \citep[][]{Oeberg2011, Helling2014, Eistrup2018}.

\begin{figure*}[htb]
\includegraphics[width=9cm]{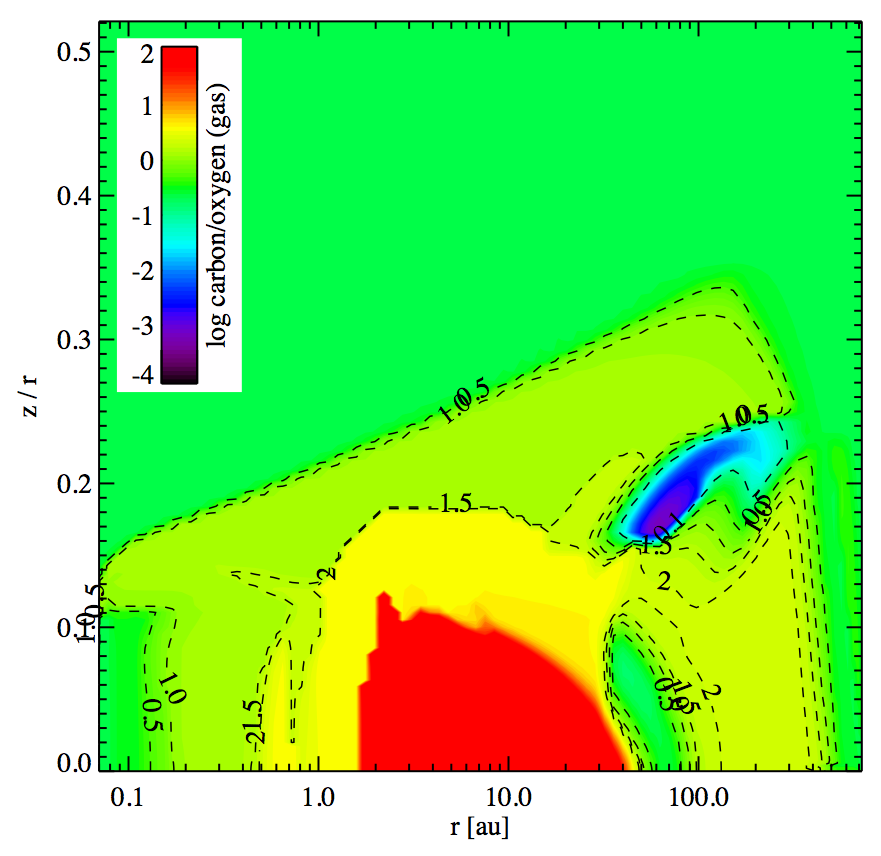}
\includegraphics[width=9cm]{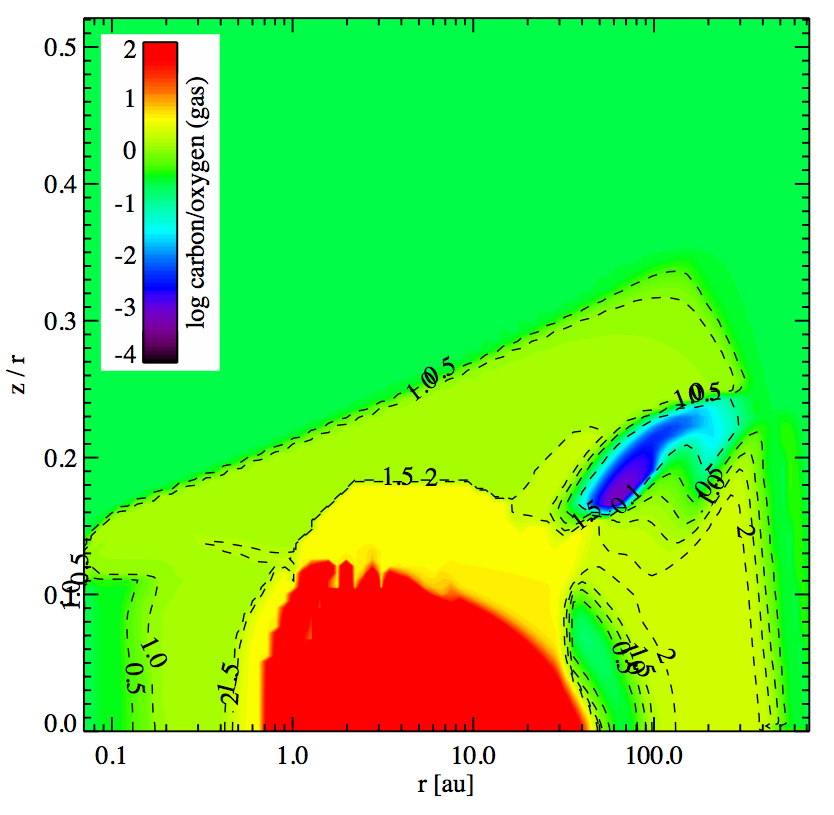}
\caption{\jm{Gas-phase C-to-O ratio} of the T Tauri reference model (left) and the model with instanton rate H$_2$~+~OH~$\rightarrow$~H$_2$O~+~H (right).}
\label{fig:TTauri_CoverO}
\end{figure*}

We compared the C-to-O ratio in the gas phase between the reference T Tauri model and the model using the values of the instanton calculations for the reaction H$_2$~+~OH~$\rightarrow$~H$_2$O~+~H [1]. 
Differences occur close to the midplane snowline (below $z/r \sim 0.1$). Due to an enhanced formation route of water in the gas phase, water ice formation is enhanced, and \jm{thus the C-to-O ratio} becomes higher than 10 already inside 1~au (Fig.~\ref{fig:TTauri_CoverO}). 
The change in water formation also affects the water ice-to-rock ratio close to the snowline. This ratio determines the enhancement in solid mass available for the formation of planetary embryos \citep[e.g.,][for planet population synthesis models]{Ida2008,Benz2014}. We do not include the formation of water on surfaces through H addition in this work, and hence the results above should be seen as differential rather than absolute.

A change in the water formation around 
the snowline can also affect the buildup of atmospheres of terrestrial planets. \citet{DAngelo2018} and \citet{Thi2018} show that the water vapor pressure inside the snowline also affects  the formation of phyllosilicates and thus the content of hydrous minerals available for the formation of planetary cores.

\section{Conclusion}

We provide Arrhenius-type low-temperature fits for rate coefficients of gas-phase reactions that have been identified as being potentially affected by tunneling at low temperature ($T$~<~250~K). For CH$_4$ reacting with OH, H, and O (reactions [2], [3], and [41], respectively) these rate constants become orders of magnitude larger compared to the extrapolated UMIST2012 rate constants. 
For the reaction 
H$_2$~+~OH~$\rightarrow$~H$_2$O~+~H, this occurs at temperatures below the validity range of 250~K given in the UMIST2012 database. The rate constant for the reaction H$_2$~+~CN~$\rightarrow$~HCN~+~H becomes a factor 10 smaller than the UMIST2012 rate constant below their low-temperature limit.
We did not identify any ion-molecule reactions that play a relevant role in disk models and that are affected by tunneling. The new Arrhenius parameters fitted to the low-temperature rate constant obtained from the literature could be included in future database releases.

An important result of this study is that in \jm{planet-forming} disks, the temperature region where atom tunneling in gas-phase chemistry could be important (approximately where $140 \lesssim T \lesssim 250$~K) is generally small. It is either limited by ice formation and hence surface chemistry taking over at low temperatures (ice reservoir) or radiation-dominated environments causing gas temperatures well above 250~K and ion-molecule chemistry to dominate (upper disk surface layers). The transition regime is either radially or vertically very thin except for special disk geometries (e.g.,\ flat disk models). 

The regions affected are often below the disk layers that give rise to the wealth of emission lines at IR to submm wavelengths. Thus, predicted changes in emitting line fluxes from key molecules such as CO, OH, H$_2$O, CN, and HCN are smaller than 20\%. However, close to the snowline, atom tunneling in one of the main water formation routes
that  affects the gas-phase C-to-O ratio and the ice-to-rock ratio, both quantities that are relevant for planet and planetary atmosphere formation. 

Future steps for \jm{planet-forming} disks should be a detailed study of the surface chemistry including tunneling effects. This is especially relevant at the interface between gas and ices where, for example,\ formation of water ice through adsorption from the gas phase can compete with surface formation of water through adsorption of H and O atoms. Surface chemistry can also enhance gas-phase abundances of species like water provided that reactive desorption is efficient. While at greater disk heights the water vapor and ice abundances are affected by surface reactions and nonthermal and reactive desorption, this process is less relevant close to the midplane around the snowline \citep[][Thi et al.\ in preparation]{Walsh2014}.
 \jm{
Other astrophysical environments such as molecular clouds, hot cores, and hot corinos should also be analyzed in detail since 
\jm{the low densities and/or time-dependence of the warm-up processes of these objects 
change the pathways of the gas-phase chemistry.}
}

The tunneling effect is mass-dependent, 
leading to a particularly pronounced kinetic isotope effect for hydrogen. Therefore, the distribution of
deuterium atoms in molecules can significantly deviate from the atomic H-to-D ratio. 
Including D atoms in the chemical network is a project by itself 
and the HDO-to-H$_2$O ratio is key to many studies on the formation of 
solar system objects.
Therefore, we aim to extend our studies presented here to 
show how decisive atom tunneling is for H-to-D ratios 
and how existing reaction networks can be extended to reliably describe deuterium transfer.

\begin{acknowledgements}
We would like to thank Laurent Wiesenfeld, Stephanie Cazaux, Thanja Lamberts, and Tom Millar for the insightful discussions. 
JM thanks the  German  Research Foundation (DFG)
for financial support within the Cluster of Excellence in Simulation 
Technology (EXC 310/2) at the University of Stuttgart.
JK thanks European  Union’s  Horizon  2020  Research  and  Innovation  Programme (Grant Agreement 646717, TUNNELCHEM). 
We thank the COST Action CM1401 ``Our Astro-Chemical History'' for travel support through two STSMs and stimulating discussions.
\end{acknowledgements}

\clearpage
\newpage

\begin{appendix}

\section{Calculated rate constants}
\label{app:CalculatedRatesPlots}

%
%
%
%

In the following we discuss the Arrhenius plots for reactions
[2], [3], [24], [26], [28], [35], and [41].
The values used for the fitting procedure are denoted by dots in the following and are given in the  cited literature.

\begin{figure}[htb]
\includegraphics[width=9cm]{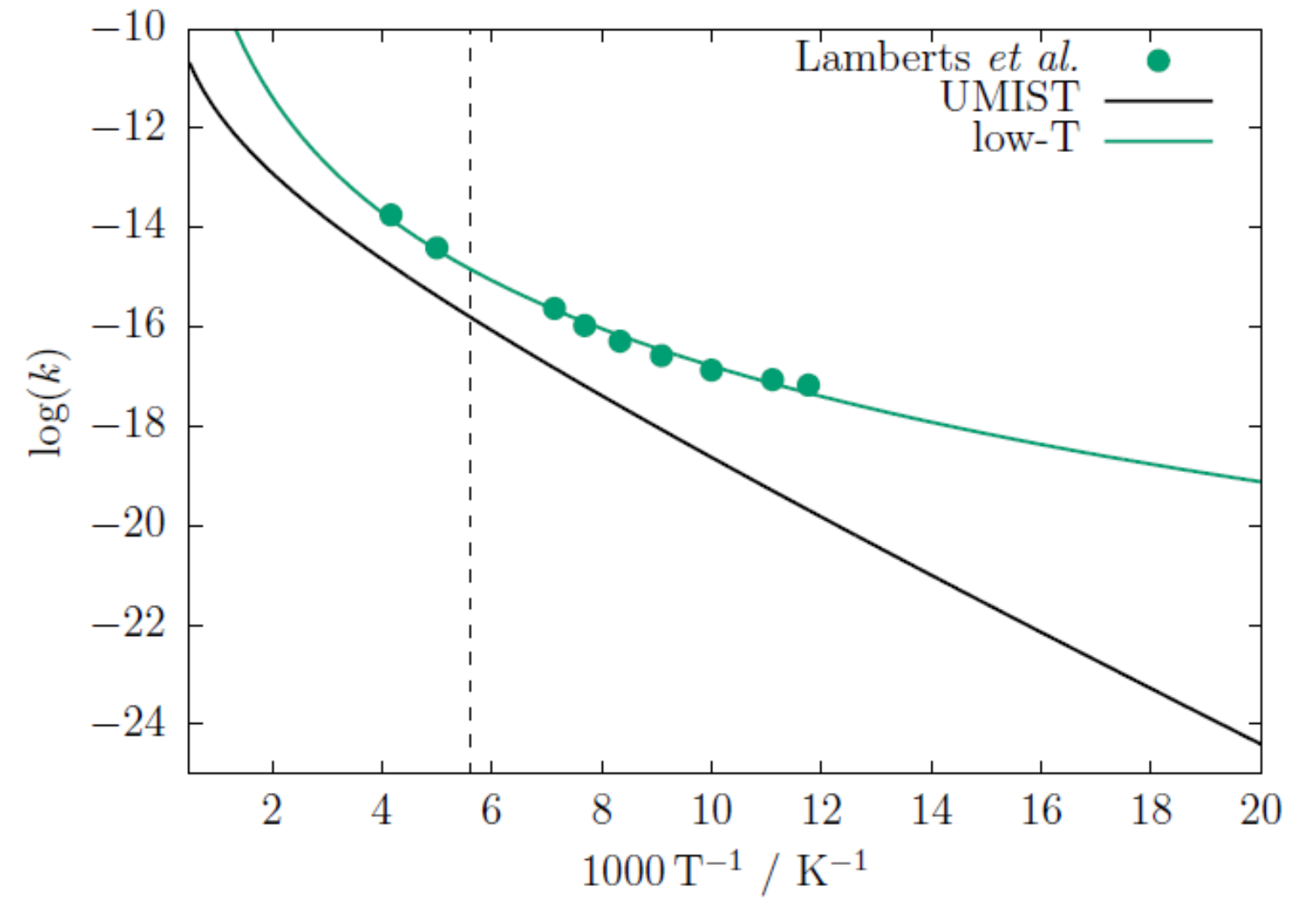}
\caption{Arrhenius plot of the rate constants using the  UMIST2012 
parameters and the fit the literature values for the reaction 
CH$_4$~+~OH$~\rightarrow$~CH$_3$~+~H$_2$O~[2]. The instanton values have been published by \citet{lamberts2017}
    The vertical dashed line indicates the lower bound for the recommended temperature range of the UMIST2012 database, $T_l =178$~K.
}
\label{fig:CH4+OH}
\end{figure}

\begin{figure}[htb]
\includegraphics[width=9cm]{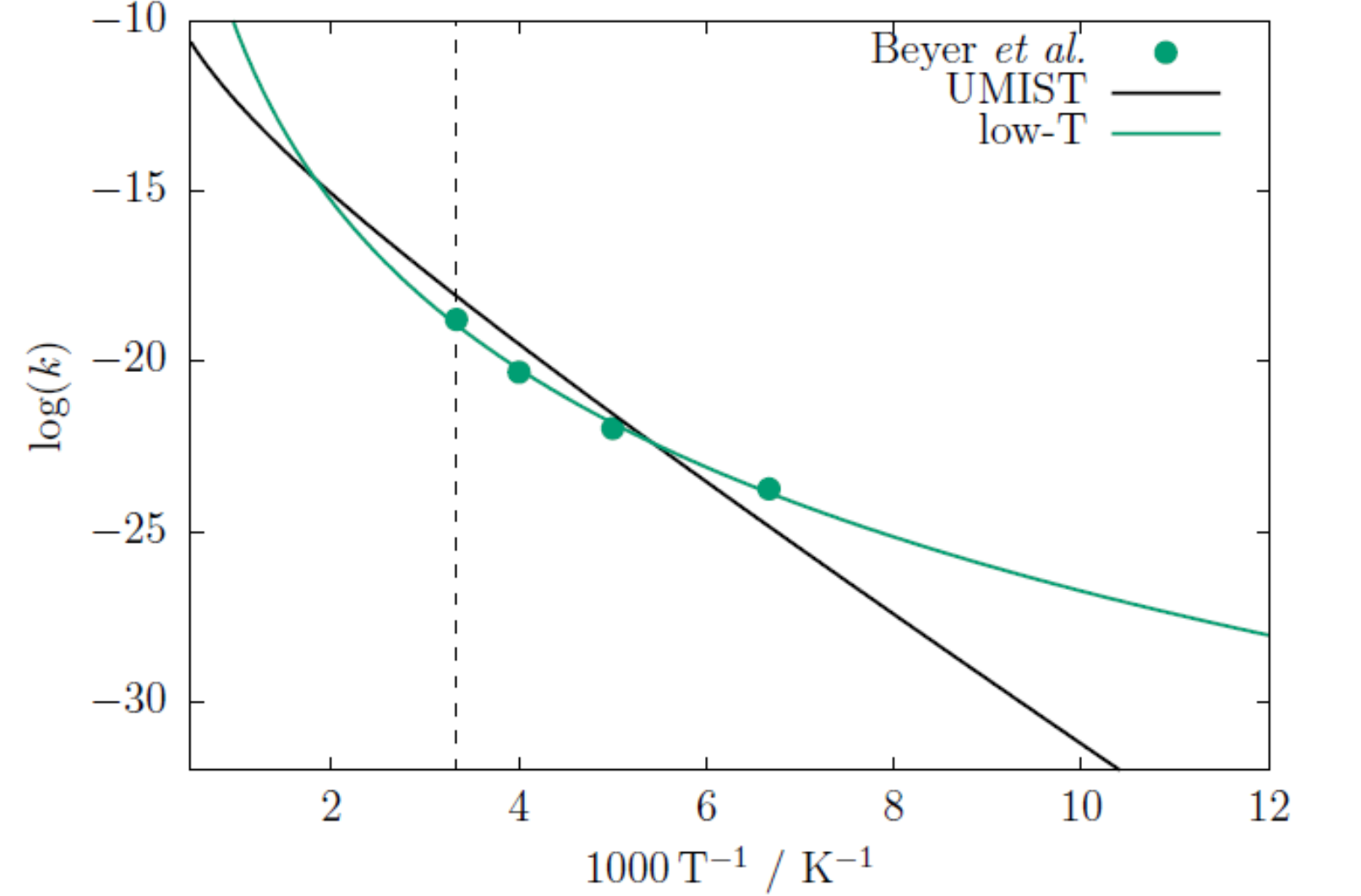}
\caption{Arrhenius plot of the rate constants using the  UMIST2012 
parameters and the fit the literature values for the reaction CH$_4$~+~H$~\rightarrow$~CH$_3$~+~H$_2$~[3]. The instanton values are from \citet{beyer2016}
   The vertical dashed line indicates the lower bound for the recommended temperature range of the UMIST2012 database, $T_l =300$~K.
}
\label{fig:CH4+H}
\end{figure}

\begin{figure}[htb]
\includegraphics[width=9cm]{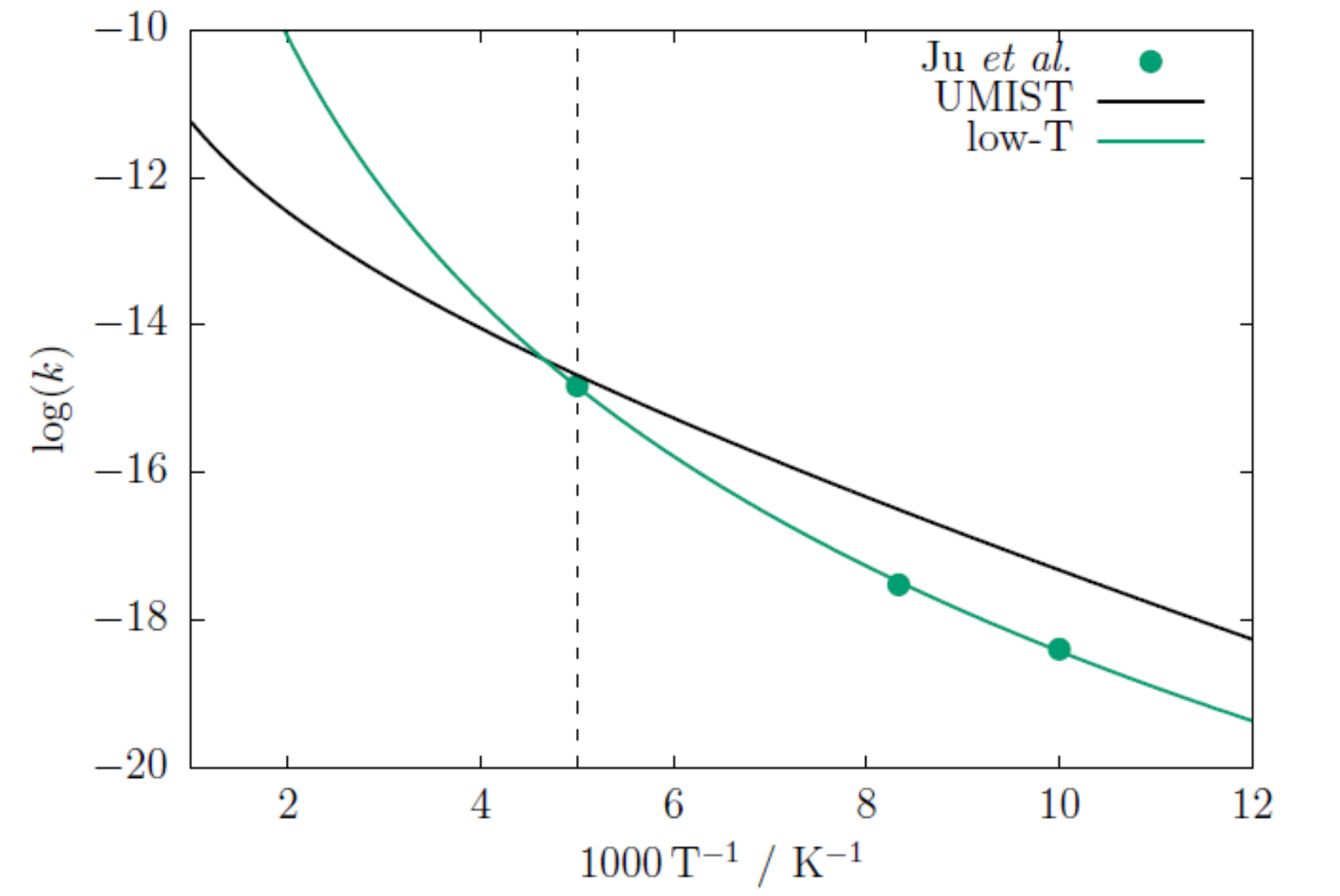}
\caption{Arrhenius plot of the rate constants using the UMIST2012 
parameters and the fit the literature values for the reaction  H$_2$~+~CN~$\rightarrow$~HCN~+~H~[24]. The values are from canonical-variational transition state theory and small-curvature Tunneling corrections (ICVT/SCT)\citet{ju2006}
     The vertical dashed line indicates the lower bound for the recommended temperature range of the UMIST2012 database,  $T_l =200$~K.
}
\label{fig:H2+CN}
\end{figure}

\begin{figure}[htb]
\includegraphics[width=9cm]{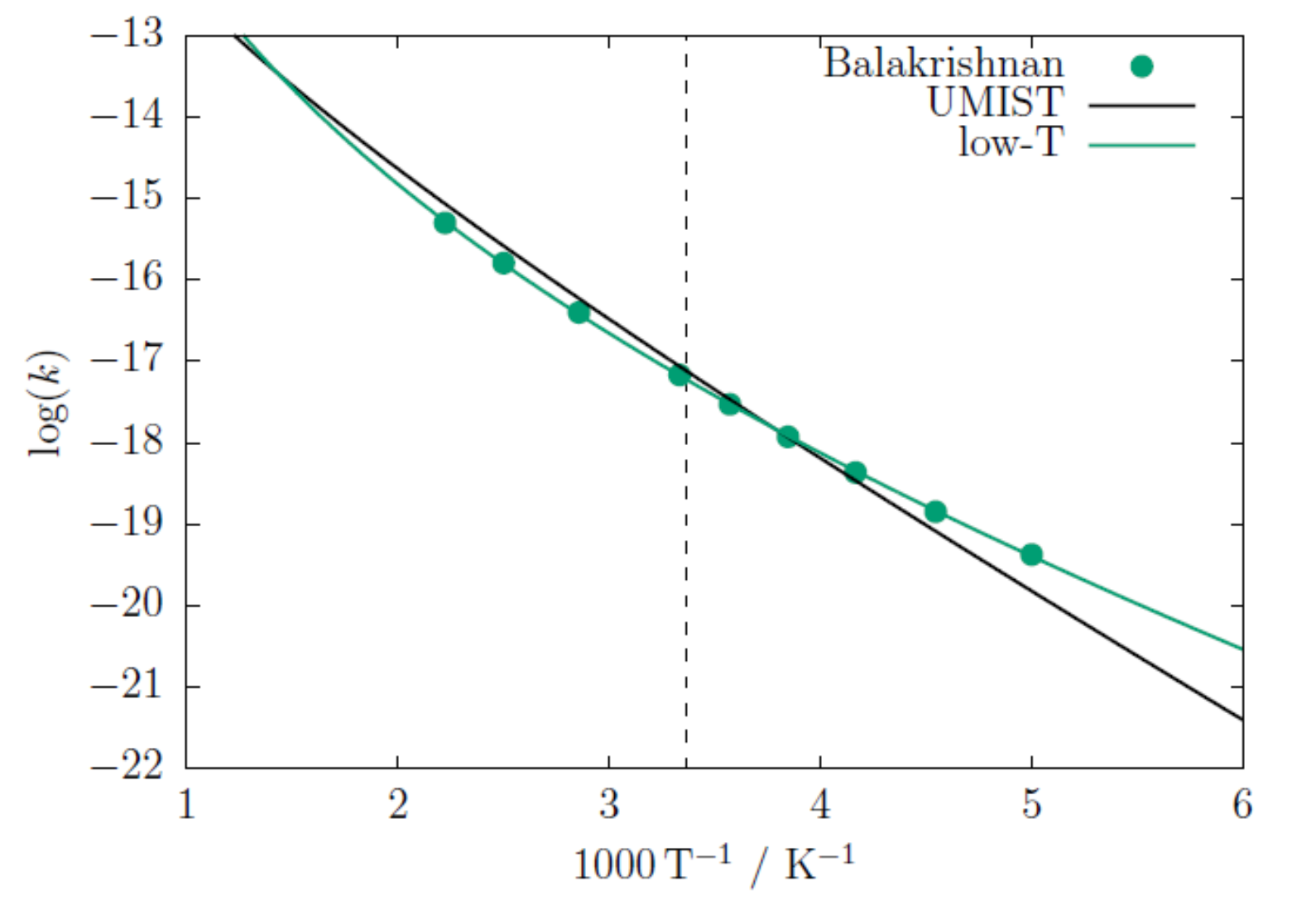}
\caption{Arrhenius plot of the rate constants using the UMIST2012 
parameters and the fit of the literature values for the reaction H$_2$~+~O$~\rightarrow$~H~+~OH~[26]. The values used for the fit are from \citet{balakrishnan2003}.
     The vertical dashed line indicates the lower bound for the recommended temperature range of the UMIST2012 database, $T_l =298$~K.
}
\label{fig:H2+O}
\end{figure}

\begin{figure}[htb]
\includegraphics[width=9cm]{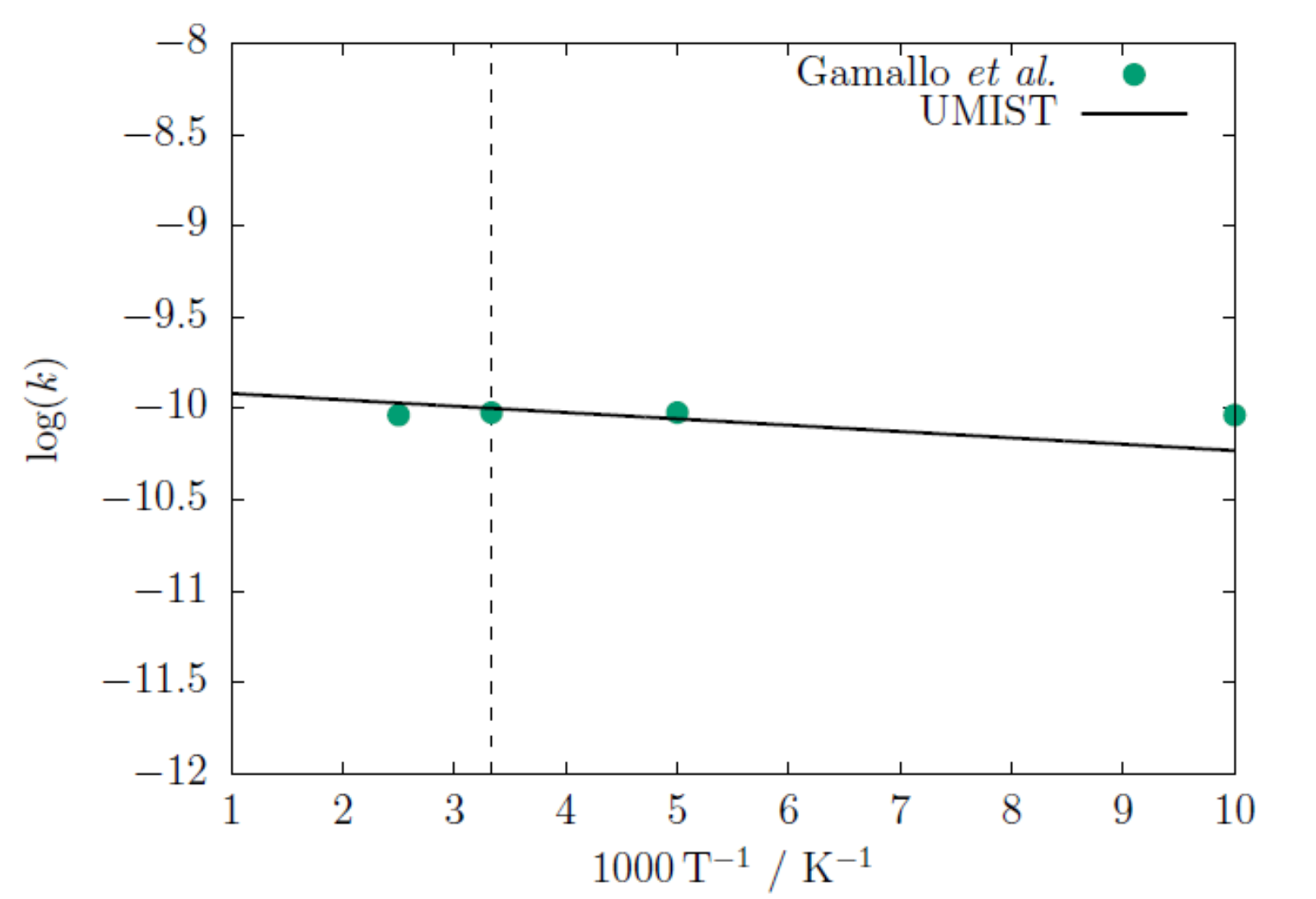}
\caption{Arrhenius plot of the rate constants using the  UMIST2012 parameters for the reaction H~+~CH~$\rightarrow$~C~+~H$_2$~[28]. 
As the values of  \citet{gamallo2012} obtained by quantum dynamics calculations agree very well with the rate constants using the UMIST2012 parameters, no fit was carried out.
     The vertical dashed line indicates the lower bound for the recommended temperature range of the UMIST2012 database,  $T_l =300$~K.
}
\label{fig:H+CH}
\end{figure}

\begin{figure}[htb]
\includegraphics[width=9cm]{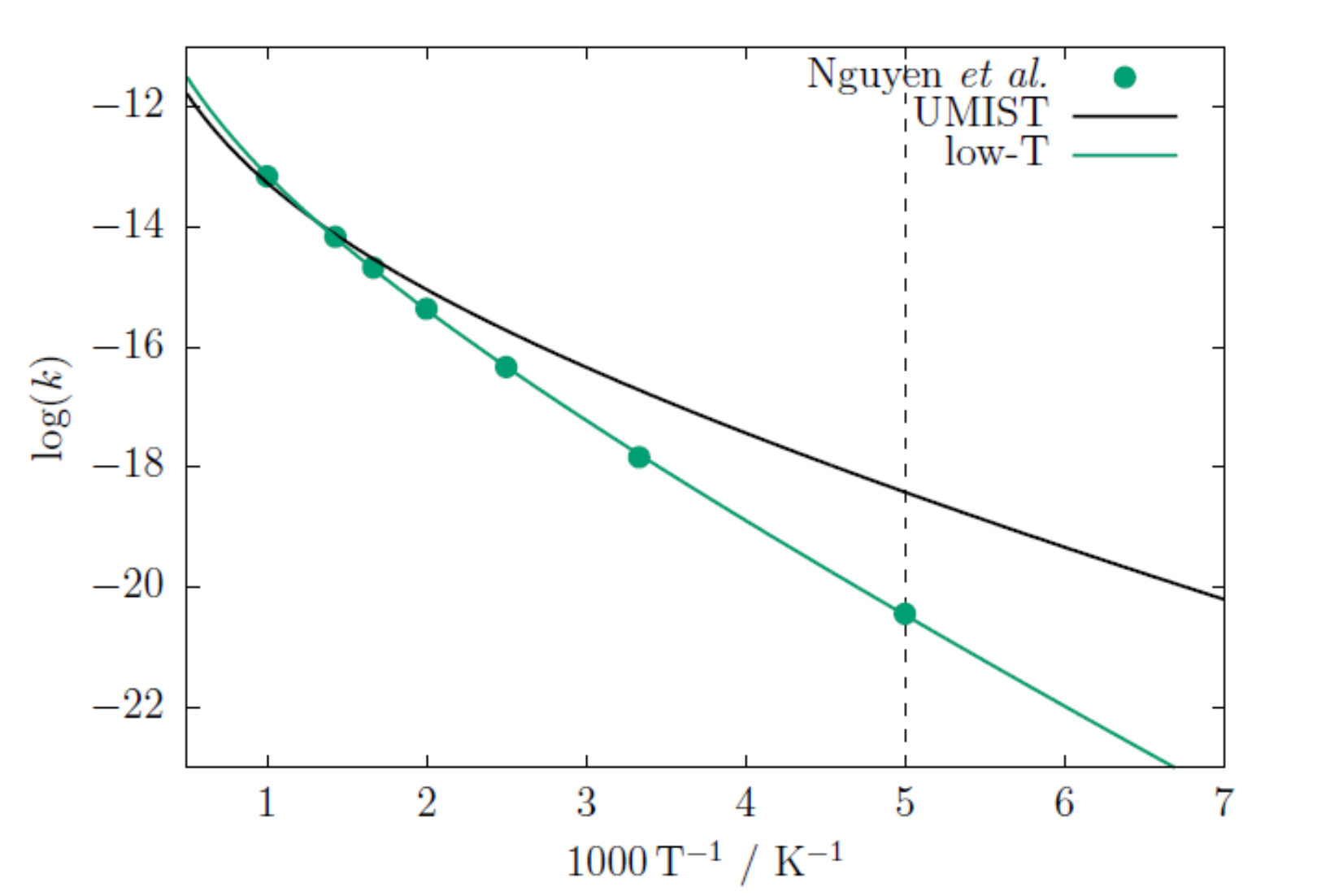}
\caption{Arrhenius plot of the rate constants using the  UMIST2012 parameters for the reaction NH$_2$~+~H$_2 \rightarrow$~NH$_3$~+~H~[35].
The computational values of \citet{nguyen2019} are noticeably below the UMIST2012 fit. 
     The vertical dashed line indicates the lower bound for the recommended temperature range of the UMIST2012 database,  $T_l =200$~K.
}
\label{fig:NH2+H2}
\end{figure}

\begin{figure}[htb]
\includegraphics[width=9cm]{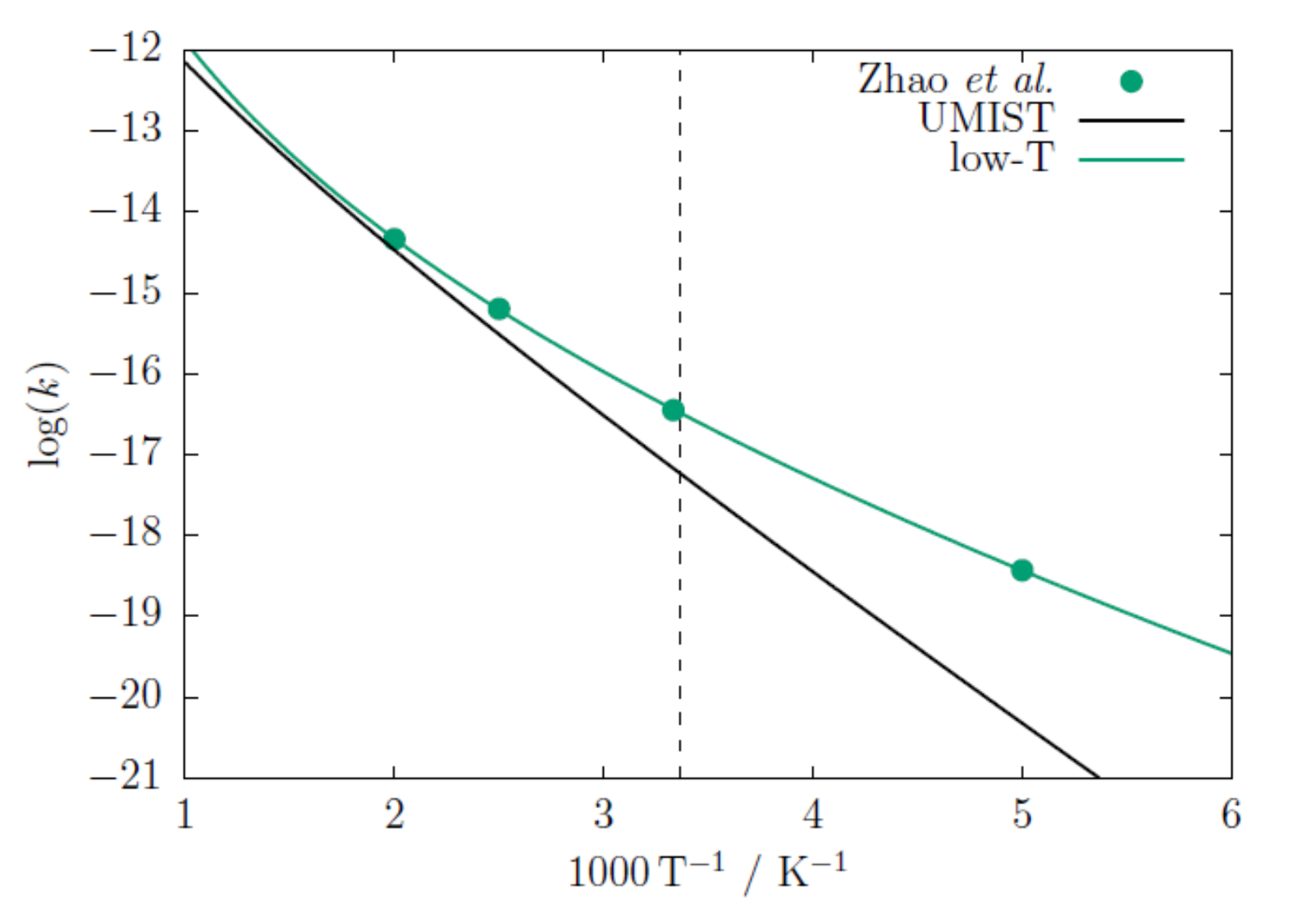}
\caption{Arrhenius plot of the rate constants using the  UMIST2012 parameters and the fit the literature values for the reaction 
O~+~CH$_4 \rightarrow$~OH~+~CH$_3$~[41].
The fit was performed using the values of \citep{zhao2016}
obtained by quantum instanton calculations.
     The vertical dashed line indicates the lower bound for the recommended temperature range of the UMIST2012 database, $T_l =298$~K.
}
\label{fig:CH4+O}
\end{figure}


\section{Line flux comparison}

\begin{figure*}[htb]
\includegraphics[width=18cm]{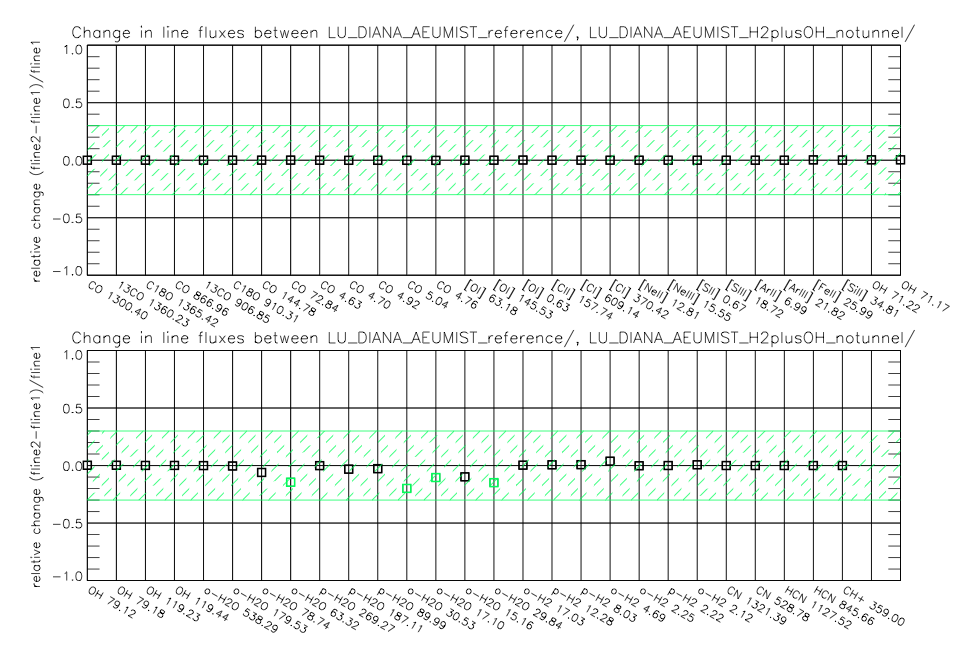}
\caption{Comparison of emission line fluxes from the T Tauri reference model and the model using
the Arrhenius parameters of the reaction H$_2$~+~OH~$\rightarrow$~H$_2$O~+~H~[1] calculated 
explicitly without atom tunneling.}
\label{fig:compare-lines-ref-notunnel}
\end{figure*}

We present here as an example a complete overview of the changes in line fluxes emergent from the disk model. The reference model is the T Tauri disk and the comparison model is  with reaction H$_2$~+~OH~$\rightarrow$~H$_2$O~+~H~[1] calculated without taking tunneling into account (Sect.~\ref{sect:no-tunnel}). Figure~\ref{fig:compare-lines-ref-notunnel} shows that some high-excitation water lines deviate by up to $\sim 60$\%.

\end{appendix}

\bibliographystyle{aa}
\bibliography{references,NEW}

\begin{thebibliography}{71}
\expandafter\ifx\csname natexlab\endcsname\relax\def\natexlab#1{#1}\fi

\bibitem[{\'{A}lvarez Barcia {et~al.}(2016)\'{A}lvarez Barcia, Russ, Meisner,
  \& K\"{a}stner}]{alvarez-barcia2016}
\'{A}lvarez Barcia, S., Russ, M.-S., Meisner, J., \& K\"{a}stner, J. 2016,
  Faraday Discuss., 195, 69

\bibitem[{Andersson {et~al.}(2011)Andersson, Goumans, \&
  Arnaldsson}]{andersson2011}
Andersson, S., Goumans, T. P.~M., \& Arnaldsson, A. 2011, Chem. Phys. Lett.,
  513, 31

\bibitem[{{Aresu} {et~al.}(2012){Aresu}, {Meijerink}, {Kamp}, {Spaans}, {Thi},
  \& {Woitke}}]{Aresu2012}
{Aresu}, G., {Meijerink}, R., {Kamp}, I., {et~al.} 2012, \aap, 547, A69

\bibitem[{Balakrishnan(2003)}]{balakrishnan2003}
Balakrishnan, N. 2003, J. Chem. Phys., 119, 195

\bibitem[{{Bensch}(2006)}]{Bensch2006}
{Bensch}, F. 2006, \aap, 448, 1043

\bibitem[{{Benz} {et~al.}(2014){Benz}, {Ida}, {Alibert}, {Lin}, \&
  {Mordasini}}]{Benz2014}
{Benz}, W., {Ida}, S., {Alibert}, Y., {Lin}, D., \& {Mordasini}, C. 2014,
  Protostars and Planets VI, 691

\bibitem[{{Bergin} {et~al.}(2010){Bergin}, {Phillips}, {Comito}, {Crockett},
  {Lis}, {Schilke}, {Wang}, {Bell}, {Blake}, {Bumble}, {Caux}, {Cabrit},
  {Ceccarelli}, {Cernicharo}, {Daniel}, {de Graauw}, {Dubernet},
  {Emprechtinger}, {Encrenaz}, {Falgarone}, {Gerin}, {Giesen}, {Goicoechea},
  {Goldsmith}, {Gupta}, {Hartogh}, {Helmich}, {Herbst}, {Joblin}, {Johnstone},
  {Kawamura}, {Langer}, {Latter}, {Lord}, {Maret}, {Martin}, {Melnick},
  {Menten}, {Morris}, {M{\"u}ller}, {Murphy}, {Neufeld}, {Ossenkopf}, {Pagani},
  {Pearson}, {P{\'e}rault}, {Plume}, {Roelfsema}, {Qin}, {Salez}, {Schlemmer},
  {Stutzki}, {Tielens}, {Trappe}, {van der Tak}, {Vastel}, {Yorke}, {Yu}, \&
  {Zmuidzinas}}]{Bergin2010}
{Bergin}, E.~A., {Phillips}, T.~G., {Comito}, C., {et~al.} 2010, \aap, 521, L20

\bibitem[{Beyer {et~al.}(2016)Beyer, Richardson, Knowles, Rommel, \&
  Althorpe}]{beyer2016}
Beyer, A.~N., Richardson, J.~O., Knowles, P.~J., Rommel, J., \& Althorpe, S.~C.
  2016, J. Phys. Chem. Lett., 7, 4374

\bibitem[{Biczysko {et~al.}(2018)Biczysko, Bloino, \& Puzzarini}]{biczysko2018}
Biczysko, M., Bloino, J., \& Puzzarini, C. 2018, Wiley Interdisciplinary
  Reviews: Computational Molecular Science, 8, e1349

\bibitem[{Borden(2016)}]{borden2016}
Borden, W.~T. 2016, WIREs Comput. Mol. Sci., 6, 20

\bibitem[{{Carr} \& {Najita}(2008)}]{Carr2008}
{Carr}, J.~S. \& {Najita}, J.~R. 2008, Science, 319, 1504

\bibitem[{{Cazaux} {et~al.}(2016){Cazaux}, {Minissale}, {Dulieu}, \&
  {Hocuk}}]{Cazaux2016}
{Cazaux}, S., {Minissale}, M., {Dulieu}, F., \& {Hocuk}, S. 2016, \aap, 585,
  A55

\bibitem[{{Cazzoletti} {et~al.}(2018){Cazzoletti}, {van Dishoeck}, {Visser},
  {Facchini}, \& {Bruderer}}]{Cazzoletti2018}
{Cazzoletti}, P., {van Dishoeck}, E.~F., {Visser}, R., {Facchini}, S., \&
  {Bruderer}, S. 2018, \aap, 609, A93

\bibitem[{{Crockett} {et~al.}(2014){Crockett}, {Bergin}, {Neill}, {Favre},
  {Schilke}, {Lis}, {Bell}, {Blake}, {Cernicharo}, {Emprechtinger},
  {Esplugues}, {Gupta}, {Kleshcheva}, {Lord}, {Marcelino}, {McGuire},
  {Pearson}, {Phillips}, {Plume}, {van der Tak}, {Tercero}, \&
  {Yu}}]{Crockett2014}
{Crockett}, N.~R., {Bergin}, E.~A., {Neill}, J.~L., {et~al.} 2014, \apj, 787,
  112

\bibitem[{{D'Angelo} {et~al.}(2018){D'Angelo}, {Cazaux}, {Kamp}, {Thi}, \&
  {Woitke}}]{DAngelo2018}
{D'Angelo}, M., {Cazaux}, S., {Kamp}, I., {Thi}, W.-F., \& {Woitke}, P. 2018,
  ArXiv e-prints

\bibitem[{{Deming} \& {Seager}(2017)}]{Deming2017}
{Deming}, L.~D. \& {Seager}, S. 2017, Journal of Geophysical Research
  (Planets), 122, 53

\bibitem[{{Eistrup} {et~al.}(2018){Eistrup}, {Walsh}, \& {van
  Dishoeck}}]{Eistrup2018}
{Eistrup}, C., {Walsh}, C., \& {van Dishoeck}, E.~F. 2018, \aap, 613, A14

\bibitem[{Gamallo {et~al.}(2012)Gamallo, Defazio, Akpinar, \&
  Petrongolo}]{gamallo2012}
Gamallo, P., Defazio, P., Akpinar, S., \& Petrongolo, C. 2012, J. Phys. Chem.
  A, 116, 8291

\bibitem[{Goumans(2011)}]{goumans2011b}
Goumans, T. P.~M. 2011, Mon. Notices Royal Astron. Soc., 413, 2615

\bibitem[{Goumans \& K\"astner(2010)}]{goumans2010}
Goumans, T. P.~M. \& K\"astner, J. 2010, Angew. Chem. Int. Ed., 49, 7350

\bibitem[{Goumans \& K\"astner(2011)}]{goumans2011a}
Goumans, T. P.~M. \& K\"astner, J. 2011, J. Phys. Chem. A, 115, 10767

\bibitem[{{Greenwood} {et~al.}(2017){Greenwood}, {Kamp}, {Waters}, {Woitke},
  {Thi}, {Rab}, {Aresu}, \& {Spaans}}]{Greenwood2017}
{Greenwood}, A.~J., {Kamp}, I., {Waters}, L.~B.~F.~M., {et~al.} 2017, \aap,
  601, A44

\bibitem[{Hama \& Watanabe(2013)}]{hama2013}
Hama, T. \& Watanabe, N. 2013, Chem. Rev., 113, 8783

\bibitem[{{Harada} {et~al.}(2010){Harada}, {Herbst}, \& {Wakelam}}]{Harada2010}
{Harada}, N., {Herbst}, E., \& {Wakelam}, V. 2010, \apj, 721, 1570

\bibitem[{{Helling} {et~al.}(2014){Helling}, {Woitke}, {Rimmer}, {Kamp}, {Thi},
  \& {Meijerink}}]{Helling2014}
{Helling}, C., {Woitke}, P., {Rimmer}, P.~B., {et~al.} 2014, Life, 4

\bibitem[{Herbst {et~al.}(1998)Herbst, DeFrees, Talbi, Pauzat, Koch, \&
  McLean}]{herbst1998}
Herbst, E., DeFrees, D.~J., Talbi, D., {et~al.} 1998, J. Chem. Phys., 94, 7842

\bibitem[{Hidaka {et~al.}(2009)Hidaka, Watanabe, Kouchi, \&
  Watanabe}]{hidaka2009}
Hidaka, H., Watanabe, M., Kouchi, A., \& Watanabe, N. 2009, Astrophys. J., 702,
  291

\bibitem[{{Ida} \& {Lin}(2008)}]{Ida2008}
{Ida}, S. \& {Lin}, D.~N.~C. 2008, \apj, 685, 584

\bibitem[{{Isella} {et~al.}(2007){Isella}, {Testi}, {Natta}, {Neri}, {Wilner},
  \& {Qi}}]{Isella2007}
{Isella}, A., {Testi}, L., {Natta}, A., {et~al.} 2007, \aap, 469, 213

\bibitem[{Ju {et~al.}(2006)Ju, Han, \& Zhang}]{ju2006}
Ju, L.-P., Han, K.-L., \& Zhang, J. Z.~H. 2006, J. Chem. Theory Comput., 05,
  769

\bibitem[{{Kamp} {et~al.}(2013){Kamp}, {Thi}, {Meeus}, {Woitke}, {Pinte},
  {Meijerink}, {Spaans}, {Pascucci}, {Aresu}, \& {Dent}}]{Kamp2013}
{Kamp}, I., {Thi}, W.-F., {Meeus}, G., {et~al.} 2013, \aap, 559, A24

\bibitem[{{Kamp} {et~al.}(2017){Kamp}, {Thi}, {Woitke}, {Rab}, {Bouma}, \&
  {M{\'e}nard}}]{Kamp2017}
{Kamp}, I., {Thi}, W.-F., {Woitke}, P., {et~al.} 2017, \aap, 607, A41

\bibitem[{{Kamp} {et~al.}(2010){Kamp}, {Tilling}, {Woitke}, {Thi}, \&
  {Hogerheijde}}]{Kamp2010}
{Kamp}, I., {Tilling}, I., {Woitke}, P., {Thi}, W., \& {Hogerheijde}, M. 2010,
  \aap, 510, A260000+

\bibitem[{Lamberts {et~al.}(2017)Lamberts, Fedoseev, K\"{a}stner, Ioppolo, \&
  Linnartz}]{lamberts2017}
Lamberts, T., Fedoseev, G., K\"{a}stner, J., Ioppolo, S., \& Linnartz, H. 2017,
  Astron. Astrophys., 599, A132

\bibitem[{{Line} {et~al.}(2014){Line}, {Knutson}, {Wolf}, \& {Yung}}]{Line2014}
{Line}, M.~R., {Knutson}, H., {Wolf}, A.~S., \& {Yung}, Y.~L. 2014, \apj, 783,
  70

\bibitem[{{Madhusudhan} {et~al.}(2017){Madhusudhan}, {Bitsch}, {Johansen}, \&
  {Eriksson}}]{Madhusudhan2017}
{Madhusudhan}, N., {Bitsch}, B., {Johansen}, A., \& {Eriksson}, L. 2017,
  \mnras, 469, 4102

\bibitem[{{McElroy} {et~al.}(2013){McElroy}, {Walsh}, {Markwick}, {Cordiner},
  {Smith}, \& {Millar}}]{McElroy2013b}
{McElroy}, D., {Walsh}, C., {Markwick}, A.~J., {et~al.} 2013, \aap, 550, A36

\bibitem[{Meisner(2018)}]{meisnerthesis2018}
Meisner, J. 2018, PhD thesis, University of Stuttgart

\bibitem[{Meisner \& K\"astner(2016{\natexlab{a}})}]{meisnerreview2016}
Meisner, J. \& K\"astner, J. 2016{\natexlab{a}}, Angew. Chem. Int. Ed., 55,
  5400

\bibitem[{Meisner \& K\"astner(2016{\natexlab{b}})}]{meisner2016}
Meisner, J. \& K\"astner, J. 2016{\natexlab{b}}, J. Chem. Phys., 144, 174303

\bibitem[{Meisner {et~al.}(2017)Meisner, Lamberts, \&
  K\"{a}stner}]{meisner2017}
Meisner, J., Lamberts, T., \& K\"{a}stner, J. 2017, ACS Earth Space Chem., 1,
  399

\bibitem[{{Min} {et~al.}(2016){Min}, {Rab}, {Woitke}, {Dominik}, \&
  {M{\'e}nard}}]{Min2016a}
{Min}, M., {Rab}, C., {Woitke}, P., {Dominik}, C., \& {M{\'e}nard}, F. 2016,
  \aap, 585, A13

\bibitem[{{Najita} {et~al.}(2011){Najita}, {{\'A}d{\'a}mkovics}, \&
  {Glassgold}}]{Najita2011}
{Najita}, J.~R., {{\'A}d{\'a}mkovics}, M., \& {Glassgold}, A.~E. 2011, \apj,
  743, 147

\bibitem[{Nguyen \& Stanton(2019)}]{nguyen2019}
Nguyen, T.~L. \& Stanton, J.~F. 2019, International Journal of Chemical
  Kinetics, 0

\bibitem[{Oba {et~al.}(2012)Oba, Watanabe, Hama, Kuwahata, Hidaka, \&
  Kouchi}]{oba2012}
Oba, Y., Watanabe, N., Hama, T., {et~al.} 2012, Astrophys. J., 749, 67

\bibitem[{{{\"O}berg} {et~al.}(2011){{\"O}berg}, {Murray-Clay}, \&
  {Bergin}}]{Oeberg2011}
{{\"O}berg}, K.~I., {Murray-Clay}, R., \& {Bergin}, E.~A. 2011, \apjl, 743, L16

\bibitem[{Orkin {et~al.}(2006)Orkin, Kozlov, Poskrebyshev, \&
  Kurylo}]{orkin2006}
Orkin, V.~L., Kozlov, S.~N., Poskrebyshev, G.~A., \& Kurylo, M.~J. 2006, J.
  Phys. Chem. A, 110, 6978

\bibitem[{{Pontoppidan} {et~al.}(2010){Pontoppidan}, {Salyk}, {Blake},
  {Meijerink}, {Carr}, \& {Najita}}]{Pontoppidan2010}
{Pontoppidan}, K.~M., {Salyk}, C., {Blake}, G.~A., {et~al.} 2010, \apj, 720,
  887

\bibitem[{Ravishankara {et~al.}(1981)Ravishankara, Nicovich, Thompson, \&
  Tully}]{ravishankara1981}
Ravishankara, A.~R., Nicovich, J.~M., Thompson, R.~L., \& Tully, F.~P. 1981, J.
  Phys. Chem., 85, 2498

\bibitem[{Shannon {et~al.}(2013)Shannon, Blitz, Goddard, \&
  Heard}]{shannon2013}
Shannon, R.~J., Blitz, M.~A., Goddard, A., \& Heard, D.~E. 2013, Nat. Chem., 5,
  745

\bibitem[{Smith \& Adams(1981)}]{smith1991}
Smith, D. \& Adams, N.~G. 1981, Monthly Notices of the Royal Astronomical
  Society, 197, 377

\bibitem[{Talukdar {et~al.}(1996)Talukdar, Gierczak, Goldfarb, Rudich, Rao, \&
  Ravishankara}]{talukdar1996}
Talukdar, R.~K., Gierczak, T., Goldfarb, L., {et~al.} 1996, J. Phys. Chem.,
  100, 3037

\bibitem[{{Thi} {et~al.}(2018){Thi}, {Hocuk}, {Kamp}, {Woitke}, {Rab},
  {Cazaux}, {Caselli}, \& {D'Angelo}}]{Thi2018}
{Thi}, W.~F., {Hocuk}, S., {Kamp}, I., {et~al.} 2018, arXiv e-prints

\bibitem[{{Thi} {et~al.}(2011){Thi}, {Woitke}, \& {Kamp}}]{Thi2011a}
{Thi}, W.-F., {Woitke}, P., \& {Kamp}, I. 2011, \mnras, 412, 711

\bibitem[{{Tilling} {et~al.}(2012){Tilling}, {Woitke}, {Meeus}, {Mora},
  {Montesinos}, {Riviere-Marichalar}, {Eiroa}, {Thi}, {Isella}, {Roberge},
  {Martin-Zaidi}, {Kamp}, {Pinte}, {Sandell}, {Vacca}, {M{\'e}nard},
  {Mendigut{\'{\i}}a}, {Duch{\^e}ne}, {Dent}, {Aresu}, {Meijerink}, \&
  {Spaans}}]{Tilling2012}
{Tilling}, I., {Woitke}, P., {Meeus}, G., {et~al.} 2012, \aap, 538, A20

\bibitem[{Tolman(1920)}]{tolman1920}
Tolman, R.~C. 1920, J. Am. Chem. Soc., 42, 2506

\bibitem[{Truhlar(1978)}]{truhlar1978}
Truhlar, D.~G. 1978, J. Chem. Educ., 55, 309

\bibitem[{Truhlar \& Kohen(2000)}]{truhlar2000}
Truhlar, D.~G. \& Kohen, A. 2000, Proc. Natl. Acad. Sci. U.S.A., 94, 848

\bibitem[{{van Dishoeck} {et~al.}(2008){van Dishoeck}, {Jonkheid}, \& {van
  Hemert}}]{vanDishoeck2008}
{van Dishoeck}, E.~F., {Jonkheid}, B., \& {van Hemert}, M.~C. 2008, ArXiv
  e-prints

\bibitem[{{van Zadelhoff} {et~al.}(2003){van Zadelhoff}, {Aikawa},
  {Hogerheijde}, \& {van Dishoeck}}]{vanZadelhoff2003}
{van Zadelhoff}, G.-J., {Aikawa}, Y., {Hogerheijde}, M.~R., \& {van Dishoeck},
  E.~F. 2003, \aap, 397, 789

\bibitem[{Wakelam {et~al.}(2017)Wakelam, Bron, Cazaux, Dulieu, Gry, Guillard,
  Habart, Hornek$\ae$r, Morisset, Nyman, Pirronello, Price, Valdivia, Vidali,
  \& Watanabe}]{wakelam2017}
Wakelam, V., Bron, E., Cazaux, S., {et~al.} 2017, Molecular Astrophysics, 9, 1

\bibitem[{{Wakelam} {et~al.}(2012){Wakelam}, {Herbst}, {Loison}, {Smith},
  {Chandrasekaran}, {Pavone}, {Adams}, {Bacchus-Montabonel}, {Bergeat},
  {B{\'e}roff}, {Bierbaum}, {Chabot}, {Dalgarno}, {van Dishoeck}, {Faure},
  {Geppert}, {Gerlich}, {Galli}, {H{\'e}brard}, {Hersant}, {Hickson},
  {Honvault}, {Klippenstein}, {Le Picard}, {Nyman}, {Pernot}, {Schlemmer},
  {Selsis}, {Sims}, {Talbi}, {Tennyson}, {Troe}, {Wester}, \&
  {Wiesenfeld}}]{Wakelam2012}
{Wakelam}, V., {Herbst}, E., {Loison}, J.-C., {et~al.} 2012, \apjs, 199, 21

\bibitem[{{Walsh} {et~al.}(2014){Walsh}, {Millar}, {Nomura}, {Herbst}, {Widicus
  Weaver}, {Aikawa}, {Laas}, \& {Vasyunin}}]{Walsh2014}
{Walsh}, C., {Millar}, T.~J., {Nomura}, H., {et~al.} 2014, \aap, 563, A33

\bibitem[{{Wiesenfeld} {et~al.}(2016){Wiesenfeld}, {Thi}, {Caselli}, {Faure},
  {Bizzocchi}, {Brand{\~a}o}, {Duflot}, {Herbst}, {Klippenstein},
  {Komatsuzaki}, {Puzzarini}, {Roncero}, {Teramoto}, {Toda}, {van der Avoird},
  \& {Waalkens}}]{Wiesenfeld2016}
{Wiesenfeld}, L., {Thi}, W.-F., {Caselli}, P., {et~al.} 2016, ArXiv e-prints

\bibitem[{Williams {et~al.}(2013)Williams, Kelley, \& {many others}}]{gnuplot}
Williams, T., Kelley, C., \& {many others}. 2013, Gnuplot 5.0: an interactive
  plotting program, \url{http://gnuplot.sourceforge.net/}

\bibitem[{{Woitke} {et~al.}(2009{\natexlab{a}}){Woitke}, {Kamp}, \&
  {Thi}}]{Woitke2009}
{Woitke}, P., {Kamp}, I., \& {Thi}, W.-F. 2009{\natexlab{a}}, \aap, 501, 383

\bibitem[{{Woitke} {et~al.}(2016){Woitke}, {Min}, {Pinte}, {Thi}, {Kamp},
  {Rab}, {Anthonioz}, {Antonellini}, {Baldovin-Saavedra}, {Carmona}, {Dominik},
  {Dionatos}, {Greaves}, {G{\"u}del}, {Ilee}, {Liebhart}, {M{\'e}nard},
  {Rigon}, {Waters}, {Aresu}, {Meijerink}, \& {Spaans}}]{Woitke2016}
{Woitke}, P., {Min}, M., {Pinte}, C., {et~al.} 2016, \aap, 586, A103

\bibitem[{{Woitke} {et~al.}(2009{\natexlab{b}}){Woitke}, {Thi}, {Kamp}, \&
  {Hogerheijde}}]{Woitke2009b}
{Woitke}, P., {Thi}, W.-F., {Kamp}, I., \& {Hogerheijde}, M.~R.
  2009{\natexlab{b}}, \aap, 501, L5

\bibitem[{{Woodall} {et~al.}(2007){Woodall}, {Ag{\'u}ndez}, {Markwick-Kemper},
  \& {Millar}}]{Woodall2007}
{Woodall}, J., {Ag{\'u}ndez}, M., {Markwick-Kemper}, A.~J., \& {Millar}, T.~J.
  2007, \aap, 466, 1197

\bibitem[{Zhao {et~al.}(2016)Zhao, Wang, \& Zhao}]{zhao2016}
Zhao, H., Wang, W., \& Zhao, Y. 2016, J. Phys. Chem., 120, 7589, pMID: 27640428

\bibitem[{Zuev {et~al.}(2003)Zuev, Sheridan, Albu, Truhlar, Hrovat, \&
  Borden}]{zuev2003}
Zuev, P.~S., Sheridan, R.~S., Albu, T.~V., {et~al.} 2003, Science, 299, 867

\end{thebibliography}

\end{document}